\documentstyle[aps,manuscript,psfig]{revtex}
\begin{document}
\draft
\renewcommand{\thefootnote}{\fnsymbol{footnote}}
\begin{title}
{\bf MOLECULAR SCALE CONTACT LINE HYDRODYNAMICS \\ 
     OF IMMISCIBLE FLOWS}
\end{title}
\author{Tiezheng Qian}
\address{Department of Physics and Institute of Nano Science and Technology,\\
The Hong Kong University of Science and Technology,\\ 
Clear Water Bay, Kowloon, Hong Kong, China} 
\author{Xiao-Ping Wang}
\address{Department of Mathematics,\\
The Hong Kong University of Science and Technology,\\ 
Clear Water Bay, Kowloon, Hong Kong, China} 
\author{Ping Sheng\footnote{To whom correspondence should be addressed. 
E-mail: sheng@ust.hk}}
\address{Department of Physics and Institute of Nano Science and Technology,\\
The Hong Kong University of Science and Technology,\\ 
Clear Water Bay, Kowloon, Hong Kong, China} 
\maketitle
\begin{abstract}

From extensive molecular dynamics simulations
on immiscible two-phase flows, we find the relative slipping
between the fluids and the solid wall everywhere to follow
the generalized Navier boundary condition,
in which the amount of slipping is proportional to the sum of
tangential viscous stress and the uncompensated Young stress.
The latter arises from the deviation of the fluid-fluid interface
from its static configuration. We give a continuum formulation
of the immiscible flow hydrodynamics,
comprising the generalized Navier boundary condition,
the Navier-Stokes equation,
and the Cahn-Hilliard interfacial free energy.
Our hydrodynamic model yields interfacial and velocity profiles
matching those from the molecular dynamics simulations at
the molecular-scale vicinity of the contact line.
In particular, the behavior at high capillary numbers, leading to
the breakup of the fluid-fluid interface, is accurately predicted.

\end{abstract}
\narrowtext

\section{Introduction}
\label{intro}
Immiscible two-phase flow in the vicinity of the contact line (CL),
where the fluid-fluid interface intersects the solid wall, is a classical
problem that falls beyond the framework of conventional hydrodynamics \cite
{hua,dussan,joanny,degennes,cox,koplik,robbins,sheng,keller,blake,vinal,jacqmin}.
In particular, molecular dynamics (MD)
studies have shown relative slipping between the fluids and the wall,
in violation of the no-slip boundary condition \cite{koplik,robbins}.
There have been numerous ad-hoc models \cite{hua,sheng,blake,vinal,jacqmin}
to address this phenomenon, but none was able to give a quantitative
account of the MD slip velocity profile in the molecular-scale vicinity
of the CL. While away from the moving CL the small amount of relative slipping
was found to follow the Navier boundary condition (NBC) \cite{nbc}, i.e.,
relative slipping proportional to the tangential viscous stress,
in the molecular-scale vicinity of the CL the NBC failed totally
to account for the near-complete slip.  This failure casts doubts on the 
general applicability of the NBC to immiscible flows and
hinders a continuum formulation of the hydrodynamics in the CL region.  
In particular, a (possible) breakdown in the hydrodynamic
description for the molecular-scale CL region has been suggested
\cite{robbins}. 
In another approach \cite{hadji}, it was shown the MD results can be
reproduced by continuum finite element simulations,
provided the slip profile extracted from MD is used as input.
This work demonstrated the feasibility of the hybrid algorithm,
but left unresolved the problem concerning the boundary condition 
governing the CL motion.
Without a continuum hydrodynamic formulation,
it becomes difficult or impossible to have realistic simulations
of micro- or nanofluidics, or of immiscible flows in porous media
where the relative wetting characteristics, the moving CL dissipation,
and behavior over undulating solid surfaces may have macroscopic implications.

From MD simulations on immiscible two-phase flows, we report the finding
that the generalized Navier boundary condition (GNBC) applies for
all boundary regions, whereby the relative slipping is proportional
to the sum of tangential viscous stress and
the uncompensated Young stress.
The latter arises from the deviation of the fluid-fluid
interface from its static configuration \cite{blake}.
By combining GNBC with the Cahn-Hilliard (CH) hydrodynamic formulation of
two-phase flow \cite{vinal,jacqmin}, we obtained a consistent, continuum
description of immiscible flow with material parameters (such as
viscosity, interfacial tension, etc) directly obtainable
from MD simulations. The convective-diffusive dynamics in the vicinity
of the interface and the moving CL also means the introduction of 
two phenomenological dynamic parameters whose values can be fixed by
comparison with one MD flow profile. Once the parameter values 
are determined from MD simulations, our continuum hydrodynamics 
can yield predictions matching those from MD simulations (for different 
Couette and Poiseuille flows).
Our findings suggest the no-slip boundary condition to be
an approximation to the GNBC, accurate for most macroscopic flows
but failing in immiscible flows.
These results open the door to efficient simulations of nano- or
microfluidics involving immiscible components, as well as to 
macroscopic immiscible flow calculations, e.g., in porous media, 
that are physically meaningful at the molecular level \cite{grid}.
The latter is possible, for example, by employing the adaptive method based on
the iterative grid redistribution introduced in Ref. \cite{grid}.
This method has demonstrated the capability of resolving,
at the same time, both the global behavior of 
a partial differential equation solution with 
coarse mesh, and a strong singularity in a localized region with 
a refined local mesh of over $10^5$ ratio to the coarse mesh.

\section{Molecular dynamics simulations}
\label{md-outline}
The MD simulations were performed for
both the static and dynamic configurations in
Couette and Poiseuille flows. The two immiscible fluids
were confined between two parallel walls separated along the $z$
direction, with the fluid-solid boundaries defined by $z=0,H$
(see Fig. \ref{fig1} for Couette geometry).
Interaction between the fluid molecules was modeled by a modified
Lennard-Jones (LJ) potential
$U_{ff}=4\epsilon\left[\left(\sigma/r\right)^{12}-
\delta_{ff}\left(\sigma/r\right)^6\right]$, where $r$ is the
distance between the molecules, $\epsilon$ and $\sigma$ are the
energy scale and range of interaction, respectively, and
$\delta_{ff}=1$ for like molecules and $\delta_{ff}=-1$ for
molecules of different species. Each of the two walls was constructed
by two (or more) [001] planes of an fcc lattice 
(see Appendix \ref{app-wall-fluid}), with each wall
molecule attached to the lattice site by a harmonic spring.
The mean-squared displacement of wall molecules
was controlled to obey the Lindemann criterion.
The wall-fluid interaction was also modeled by a LJ potential
$U_{wf}$, with energy and range parameters
$\epsilon_{wf}=1.16\epsilon$ and $\sigma_{wf}=1.04\sigma$, and a
$\delta_{wf}$ for specifying the wetting property of the fluid.
Both $U_{ff}$ and $U_{wf}$ were cut off at $r_c=2.5\sigma$.
The mass of the wall molecule was set equal to that of the fluid
molecule $m$, and the average number densities for the fluids and
wall were set at $\rho=0.81/\sigma^3$ and $\rho_w=1.86/\sigma^3$,
respectively. The temperature was controlled at $2.8\epsilon/k_B$,
where $k_B$ is Boltzmann's constant.  Moving the top and bottom
walls at a constant speed $V$ in the $\pm x$ directions,
respectively, induced the Couette flow \cite{robbins}.
Applying a body force $m{\bf g}_{ext}$ to each fluid molecule in
the $x$ direction induced the Poiseuille flow \cite{koplik}. 
Periodic boundary conditions were imposed
on the $x$ and $y$ boundaries of the sample.
Most of our MD simulations were carried out on samples consisting
6144 atoms for each fluid and 2880 atoms for each wall. The sample
is $163.5\sigma$ by $6.8\sigma$ along the $x$ and $y$, respectively,
and $H=13.6\sigma$. Our MD results represent time averages over
20 to 40 million time steps. For technical details of our MD simulations, 
we followed those described in Ref. \cite{md}.

Two different cases were considered in our simulations.  The
symmetric case refers to identical wall-fluid interactions for the
two fluids (both $\delta_{wf}=1$), which leads to a flat static
interface in the $yz$ plane with a $90^\circ$ contact angle.  The
asymmetric case refers to different wall-fluid interactions, with
$\delta_{wf}=1$ for one and $\delta_{wf}=0.7$ for the other.  The
resulting static interface is a circular arc with a $64^\circ$
contact angle.  We measured six quantities in the Couette-flow
steady states of $V=0.25(\epsilon/m)^{1/2}$, $H=13.6\sigma$ for
the symmetric case and $V=0.2(\epsilon/m)^{1/2}$, $H=13.6\sigma$ for
the asymmetric case: $v_x^{slip}$, the slip velocity relative to
the moving wall; $G_x^w$, the tangential force per unit area
exerted by the wall; the $\sigma_{xx}$, $\sigma_{nx}$ components
of the fluid stress tensor ($n$ denotes the outward surface
normal), and $v_x$, $v_z$.

We denote the region within
$0.85\sigma=z_0$ of the wall the boundary layer (BL). It must be
thin enough to render sufficient precision for measuring
$v_x^{slip}$, while thick enough to fully account for
the tangential wall-fluid interaction force, due to the finite
range of the LJ interaction. Thus it is not possible to do MD 
measurements strictly at the fluid-solid boundary, not only 
because of poor statistics, but also because of this intrinsic 
limitation. The wall force can be identified
by separating the force on each fluid molecule into wall-fluid
and fluid-fluid components. For $0<z\le z_0$ the fluid molecules
can detect the atomic structure of the wall.
When coupled with kinetic collisions with the wall molecules,
there arises a nonzero tangential wall force that varies along
the $z$ direction and saturates at $z\simeq z_0$. $G_x^w$
is the saturated total tangential wall force per unit wall area
(Fig. \ref{fig2}). In Appendix \ref{app-wall-fluid} we give
account of our MD results on both the tangential and normal
components of the wall force, plus the effect(s) of increasing
the wall thickness in our simulations from $2$ layers of wall
molecules to $4$ layers and to infinite layers (by using the
continuum approximation beyond the $4$ layers).

Spatial resolution along the $x$ and $z$ directions was achieved
by evenly dividing the sampling region into bins, each
$\Delta x=0.425\sigma$ by $\Delta z =0.85\sigma$ in size.
$v_x^{slip}$ was obtained as the time average of fluid
molecules' velocities inside the BL, measured with respect to the
moving wall; $G_x^w$  was obtained from the time
average of the total tangential wall force experienced by the
fluid molecules in the BL, divided by the bin area in the $xy$
plane; $\sigma_{xx(nx)}$ was obtained from the time averages
of the kinetic momentum transfer plus the fluid-fluid interaction forces
across the constant-$x(z)$ bin surfaces, and $v_{x(z)}$ was measured
as the time-averaged velocity component(s) within each bin.
For the contribution of intermolecular forces to the stress,
we have directly measured the fluid-fluid interaction forces across
bin surfaces instead of using the Irving-Kirkwood expression
\cite{stress}, whose validity was noted to be not justified
at the fluid-fluid or the wall-fluid interface
(see the paragraph following equation (5.15) in the above reference).
In Appendix \ref{app-md-stress} we give some details on our
MD stress measurements.
As reference quantities, we also
measured $G_x^{w0}$, $\sigma_{xx}^0$, $\sigma_{nx}^0$  in the
static ($V=0$) configuration. In addition, we measured 
in both the static and dynamic configurations the average molecular
densities $\rho_1$ and $\rho_2$ for the two fluid species in each
bin to determine the interface profile.
The shear viscosity $\eta=1.95\sqrt{\epsilon
m}/\sigma^2$
and the interfacial tension $\gamma=5.5\epsilon/\sigma^2$ 
were also determined.

We have also measured the interface and velocity profiles
for the Poiseuille flow in the asymmetric case, as well as for
Couette flows with different $V$ and $H$ in the symmetric case.

\section{Generalized Navier boundary condition}
\label{md-evidence}
In the presence of a fluid-fluid interface,
the static fluid stress tensor ${\mbox{\boldmath$\sigma$}}^0$ 
contains static Young stress (surface tension) 
as well as those stresses arising from wall-fluid 
interaction. For the consideration of moving CL, we will be 
concerned with the part of the fluid stress tensor which is purely dynamic
in origin, i.e., arising purely from the hydrodynamic motion of the fluid 
(and the CL). In the notations below,
the over tilde will denote the quantity to be the difference 
between that quantity and its static part.
Thus if ${\mbox{\boldmath$\sigma$}}$ is the total stress, 
we will be concerned only with the hydrodynamic part, 
denoted by $\tilde{\mbox{\boldmath$\sigma$}}={\mbox{\boldmath$\sigma$}}
-{\mbox{\boldmath$\sigma$}}^0$.  We note that in the absence 
of body forces, force equilibrium in bulk fluid is governed by the relation
$\nabla\cdot\tilde{\mbox{\boldmath$\sigma$}}=0$. 
As shown in Fig. \ref{fig2}, this relation is altered in the BL,
where the fluid-wall interaction means the existence of a dynamic, 
tangential force density $\tilde{g}_x^w$ such that 
$\nabla\cdot\tilde{\mbox{\boldmath$\sigma$}}+\tilde{g}_x^w\hat{\bf{x}}=0$ 
inside the BL. The total tangential force exerted by the wall on the fluid 
is given by $\tilde{G}_x^w=\int_0^{z_0}dz\tilde{g}_x^w$ 
per unit wall area. In steady state, this wall force is necessarily 
balanced by the tangential fluid force $\tilde{G}_x^f=\int_0^{z_0}dz
\left(\partial_x\tilde{\sigma}_{xx}+\partial_z\tilde{\sigma}_{zx}\right)$. 
Here $\partial_{x,z,n}$ means taking partial 
derivative with respect to $x$, $z$, or surface normal.

We now present evidences to show that everywhere on the boundaries, relative
slipping is proportional to $\tilde{G}_x^f$ (the GNBC, see also equation
(\ref{gnbc3}) below):
\begin{equation}\label{gnbc1}
\tilde{G}_x^f=\beta v_x^{slip},
\end{equation}
where $\beta$  is the slip coefficient and $\tilde{G}_x^f$ can be written as
\begin{equation}\label{gnbc2}
\tilde{G}_x^f=\partial_x\int_0^{z_0}dz\tilde{\sigma}_{xx}(z)
-\tilde{\sigma}_{nx}(z_0),
\end{equation}
where we have used the fact that $\tilde{\sigma}_{zx}(0)=0$.
(More strictly, $\tilde{\sigma}_{zx}(0^-)=0$ because there is no fluid
below $z=0$ and hence no momentum transport across $z=0$)
Here the $z$ coordinate is for the lower fluid-solid boundary (same below),
with the understanding that the same physics holds at the upper boundary, 
and $\partial_n=-\partial_z$ for the lower boundary.

We have verified the steady state force balance
$\tilde{G}_x^w+\tilde{G}_x^f=0$ on the two boundaries
(inset to Fig. \ref{fig3}) \cite{balance}. This relation reflects 
the fact that at steady state, the frictional force exerted by 
the solid wall on the moving/slipping fluid is fully accounted for 
in $\tilde{G}_x^f$. Thus the GNBC (or NBC) can be expressed in either
$\tilde{G}_x^f$ or $\tilde{G}_x^w$, but not both. In Fig. \ref{fig3}
we show the measured MD data for the symmetric and asymmetric
cases in the Couette geometry. The symbols represent the values
of $\tilde{G}_x^f$ measured in the BL. The solid lines represent
the values of $\tilde{G}_x^f$ calculated from $\beta v_x^{slip}$
by using $\beta=\beta_1=\beta_2=1.2\sqrt{\epsilon m}/\sigma^3$
for the symmetric case and $\beta_1=1.2\sqrt{\epsilon m}/\sigma^3$,
$\beta_2=0.532\sqrt{\epsilon m}/\sigma^3$ for the asymmetric case
away from the CL region (straight line segments in Fig. \ref{fig3}),
and $\beta=(\beta_1\rho_1+\beta_2\rho_2)/(\rho_1+\rho_2)$
in the CL region \cite{weight}, with $v_x^{slip}$ and $\rho_{1,2}$
obtained from MD simulations. It is seen that for the lower
boundary (upper right panel), the MD data agree well with the 
predictions of Eq. (\ref{gnbc1}). For the upper boundary (lower left panel)
the straight line segments also agree well with Eq. (\ref{gnbc1}).
However, there is some discrepancy in the interfacial region of
the upper boundary that seems to arise from a ``shear thinning''
effect of decreasing $\beta$ at very large tangential stresses \cite{nbc}.

The fact that the wall force density is distributed inside a thin
BL and vanishes beyond the BL necessitates the form of
$\tilde{G}_x^f$  as defined by Eq. (\ref{gnbc2}). However, it
is intuitively obvious that the fluids would experience almost the
identical physical effect(s) from a wall force density
$\tilde{G}_x^w\delta(z)$, concentrated strictly at the fluid-solid
boundary with the same total wall force per unit area. 
In the inset to Fig. \ref{fig2}, it is shown that the MD-measured 
wall force density is a sharply peaked function. The sharp boundary limit 
involves the approximation of replacing this peaked function by $\delta(z)$.
The replacement of a diffuse boundary by a sharp boundary can
considerably simplify the form of the GNBC, because local force
balance along $x$ then requires
$\partial_x\tilde{\sigma}_{xx}+\partial_z\tilde{\sigma}_{zx}=0$
away from the boundary $z=0$. 
Integration of this relation from $0^+$ to $z_0$ yields
$\partial_x\int_0^{z_0}dz\tilde{\sigma}_{xx}(z)
+\tilde{\sigma}_{zx}(z_0)-\tilde{\sigma}_{zx}(0^+)=0$ and
as a consequence (by comparing with Eq. (\ref{gnbc2})) 
$\tilde{G}_x^f=-\tilde{\sigma}_{nx}(0^+)$. Therefore, 
$\tilde{\sigma}_{zx}$ changes from $\tilde{\sigma}_{zx}(0^-)=0$
to $\tilde{\sigma}_{zx}(0^+)=\tilde{G}_x^f$ at $z=0$, leading to
$\nabla\cdot\tilde{\mbox{\boldmath$\sigma$}}=\tilde{G}_x^f\delta(z)$.
Comparing with the diffuse boundary, we see that the form of the equation
remains the same, but the BL is now from $0^-$ to $0^+$. 
Thus the GNBC (\ref{gnbc1}) becomes
$-\tilde{\sigma}_{nx}(0)=\beta v_x^{slip}$ in the sharp boundary limit.

The tangential stress $\tilde{\sigma}_{nx}$ can be decomposed into
a viscous component and a non-viscous component:
$\tilde{\sigma}_{nx}(z)={\sigma}_{nx}^v(z)+\tilde{\sigma}_{nx}^Y(z)$.
In Fig. \ref{fig4} we show that away from the interfacial region
the tangential viscous stress ${\sigma}_{nx}^v(z) =\eta(\partial_n
v_x+\partial_x v_n)(z)$ is the only nonzero component, but in the
interfacial region 
$\tilde{\sigma}_{nx}^Y=\sigma_{nx}-\sigma_{nx}^v-\sigma_{nx}^0
=\sigma_{nx}^Y-\sigma_{nx}^0$ is dominant, thereby accounting for the
failure of NBC to describe the CL motion. Therefore away from the
CL region the NBC is valid, but in the interfacial region the NBC
clearly fails to describe the CL motion. 
We wish to clarify the origin of $\sigma_{nx}^Y$ and $\sigma_{nx}^0$
as the dynamic and static Young stresses, respectively, so that
$\tilde{\sigma}_{nx}^Y=\sigma_{nx}^Y-\sigma_{nx}^0$ is the
uncompensated Young stress. As shown in the inset to Fig. \ref{fig4},
the integrals (across the interface) of
$\sigma_{nx}^Y$ ($=\sigma_{nx}-\sigma_{nx}^v$, calculated by
subtracting the viscous component $\eta(\partial_n v_x +
\partial_x v_n)$ from the total tangential stress $\sigma_{nx}$) 
and $\sigma_{nx}^0$  are equal to 
$\gamma\cos\theta_d$ and $\gamma\cos\theta_s$,
respectively, at different values of $z$, i.e., 
$-\int_{\rm int}dx\sigma_{nx}^Y(z)=\gamma\cos\theta_d(z)$ and 
$-\int_{\rm int}dx\sigma_{nx}^0(z)=\gamma\cos\theta_s(z)$, 
where $\theta_d(z)$ and $\theta_s(z)$ are respectively the 
dynamic and static interfacial angles at $z$ \cite{tension}.
Here $\int_{\rm int}dx$ denotes the integration across
the fluid-fluid interface along $x$.
These results clearly show the origin
of the extra tangential stress in the interfacial region to be
the interfacial (uncompensated) Young stress. Thus the GNBC
is given by
\begin{equation}\label{gnbc3}
\beta v_x^{slip}=-\tilde{\sigma}_{nx}(0)=-\left[\eta\partial_n
v_x\right](0) -\tilde{\sigma}_{nx}^Y(0).
\end{equation}
Here only one component of the viscous stress is nonzero,
due to $v_n=0$ at the boundary;
and $-\tilde{\sigma}_{nx}^Y(0)$ is denoted the
uncompensated Young stress, satisfying $-\int_{\rm
int}\tilde{\sigma}_{nx}^Y(0)dx=\gamma(\cos\theta_d^{surf}-\cos\theta_s^{surf})$,
with $\theta_{d(s)}^{surf}$  being a microscopic dynamic (static)
contact angle at the fluid-solid boundary. 
The fact that $\tilde{\sigma}_{nx}^Y(0)\approx 0$ away from the CL shows
that the GNBC implies NBC for single phase flows.

Due to the diffuse nature of the BL in MD simulations, 
the contact angle $\theta_{d(s)}^{surf}$ cannot be directly measured.
Nevertheless, they are obtainable through extrapolation by
using the integrated interfacial curvature within the BL.
That is, in the sharp boundary limit the force balance in the fluids
is expressed by 
$\partial_x\tilde{\sigma}_{xx}+\partial_n\tilde{\sigma}_{nx}=0$.
Integration in $z$ across the BL gives
\begin{equation}\label{angle1}
\partial_x\int_0^{z_0}dz\tilde{\sigma}_{xx}(z)
-{\sigma}_{nx}^v(z_0)+{\sigma}_{nx}^v(0)
-\tilde{\sigma}_{nx}^Y(z_0)+\tilde{\sigma}_{nx}^Y(0)=0.
\end{equation}
Integration (of Eq. (\ref{angle1}) along $x$) across the fluid-fluid interface
then yields
\begin{equation}\label{angle2}
\Delta\left[\int_0^{z_0}dz\tilde{\sigma}_{xx}(z)\right]
-\int_{\rm int}dx{\sigma}_{nx}^v(z_0)+\int_{\rm int}dx{\sigma}_{nx}^v(0)
+\gamma{\cal K}_d-\gamma{\cal K}_s=0.
\end{equation}
where $\Delta\left[\int_0^{z_0}dz\tilde{\sigma}_{xx}(z)\right]$ 
is the change of the $z$-integrated $\tilde{\sigma}_{xx}$ across the interface,
${\cal K}_d$ and ${\cal K}_s$ denote the dynamic and static $z$-integrated 
interfacial curvatures:
$${\cal K}_d=\cos\theta_d(z_0)-\cos\theta_d^{surf},$$
and 
$${\cal K}_s=\cos\theta_s(z_0)-\cos\theta_s^{surf}.$$ 
Here $\Delta\left[\int_0^{z_0}dz\tilde{\sigma}_{xx}(z)\right]$,
${\sigma}_{nx}^v(z_0)$, $\theta_d(z_0)$, and $\theta_s(z_0)$
are obtainable from MD simulations, ${\cal K}_s\simeq\pm 2z_0\cos\theta_s^{surf}/H$ 
for the circular static interfaces, while
${\sigma}_{nx}^v(0)=\eta[\partial_n v_x](0)$ may be obtained by extrapolating
from the values of tangential viscous stress at $z=z_0$, $2z_0$, and $3z_0$. 
Therefore the microscopic dynamic contact angle $\theta_d^{surf}$ 
can be obtained from Eq. (\ref{angle2}). 
In Appendix \ref{md-continuum-correspondence} we give a more detailed account 
of the relationship between the MD measured stresses and the stress components 
in the continuum hydrodynamics. The above extrapolation is based on 
this correspondence.

We have measured the $z$-integrated
$\tilde{\sigma}_{xx}={\sigma}_{xx}-{\sigma}_{xx}^0$ in the BL.
The dominant behavior is a sharp drop across the interface, as
shown in Fig. \ref{fig5} for both the symmetric and asymmetric cases.
The value of $\theta_d^{surf}$ obtained is $88\pm 0.5^\circ$ for the symmetric 
case and $63\pm 0.5^\circ$ for the asymmetric case at the lower
boundary, and $64.5\pm 0.5^\circ$ at the upper boundary. These values are 
noted to be very close to $\theta_s^{surf}$. Yet the small difference between
the dynamic and static (microscopic) contact angles is essential 
in accounting for the near complete slip in the CL region.

In essence, our results show that in the vicinity of the CL,
the tangential viscous stress $-\sigma_{nx}^v$ as postulated by the NBC
can not give rise to the near-complete CL slip without
taking into account the tangential Young stress $-\sigma_{nx}^Y$
in combination with the gradient of the (BL-integrated) normal stress
$\sigma_{xx}$. For the static configuration, the normal stress gradient
is balanced by the Young stress, leading to the Young's equation. It is only
for a moving CL that there is a component of the Young stress
which is no longer balanced by the normal stress gradient,
and this uncompensated Young stress is precisely the additional component
captured by the GNBC but missed by the NBC.

\section{Continuum hydrodynamic formulation}
\label{model}
For Eq. (\ref{gnbc3}) to serve as a boundary condition in
hydrodynamic calculations, we need to derive the local value of the
uncompensated Young stress $\tilde{\sigma}_{nx}^Y(0)$ from a continuum
formulation of the immiscible flow hydrodynamics.
Such a formulation is important for studying the macroscopic implications
of moving CL's under scenarios beyond the capability of MD simulations.
As a first-order approximation, we formulate a hydrodynamic
model based on the GNBC and the CH free energy functional \cite{ch} that
has been successful in the calculations of fluid-fluid interfacial phenomena:
\begin{equation}\label{ch}
F[\phi]=\int d{\bf r}
\left[\displaystyle\frac{1}{2}K\left(\nabla\phi\right)^2+f(\phi)\right],
\end{equation}
where $\phi=(\rho_2-\rho_1)/(\rho_2+\rho_1)$,
$f(\phi)=-\frac{1}{2}r\phi^2+\frac{1}{4}u\phi^4$, and $K$, $r$, $u$
are parameters which can be directly obtained from MD simulations through
the interface profile thickness $\xi=\sqrt{K/r}$ \cite{interface},
the interfacial tension $\gamma=2\sqrt{2}r^2\xi/3u$, and the two
homogeneous equilibrium phases given by the condition of
$\partial f/\partial\phi=0$, yielding $\phi_\pm=\pm\sqrt{r/u}$ ($=\pm 1$
in our case).

To derive the effects of the CH free energy $F$ on immiscible flow
hydrodynamics, let us consider a composition field $\phi({\bf r})$.
A displacement of the molecules from $\bf r$  to 
${\bf r}'={\bf r}+{\bf u}({\bf r})$ induces a local change of 
$\phi$, $\delta\phi=-{\bf u}\cdot\nabla\phi$, to the first order in
$\bf u$. The associated change in $F$ is given by
the sum of a body term and a surface term:
\begin{equation}\label{ch-force}
\delta F=-\int d{\bf r}\left[{\bf g}\cdot{\bf u}\right] +\int
ds\left[\sigma_{ni}^Yu_i\right], 
\end{equation}
where ${\bf g}=\mu\nabla\phi$ is
the capillary force density, with 
$\mu=\delta F/\delta\phi=-K\nabla^2\phi-r\phi+u\phi^3$ 
being the chemical potential, and
\begin{equation}\label{young-stress}
\sigma_{ni}^Y=-K\partial_n\phi\partial_i\phi,
\end{equation} 
is
the tangential Young stress due to the spatial variation of $\phi$ 
at the fluid-solid boundary ($i\bot n$). 
Hence the two coupled equations of motion are the Navier-Stokes equation
(with the addition of the capillary force density) and 
the convection-diffusion equation for $\phi({\bf r})$:
\begin{equation}\label{he1}
\rho_m\left[{\partial{\bf v}\over
\partial t}+ \left({\bf v}\cdot\nabla\right){\bf v} \right]=
-\nabla p +\nabla\cdot{\mbox{\boldmath$\sigma$}}^v
+\mu\nabla\phi+\rho_m{\bf g}_{ext},
\end{equation}
\begin{equation}\label{he2}
{\partial\phi\over\partial t}+{\bf v}\cdot\nabla\phi=M\nabla^2\mu,
\end{equation}
together with the incompressibility condition $\nabla\cdot{\bf
v}=0$. Here $\rho_m$ is the fluid mass density, $p$ is the
pressure,
${\mbox{\boldmath$\sigma$}}^v$ denotes the viscous part of the stress
tensor, $\rho_m{\bf g}_{ext}$ is the external body force density
(for Poiseuille flows), and $M$ is the phenomenological mobility
coefficient.

Four boundary conditions are required to solve Eqs. (\ref{he1})
and (\ref{he2}). Two are given by the impermeability condition,
i.e., the normal components of the fluid velocity and diffusive flux
are zero: $v_n=0$ and $\partial_n \mu=0$. 
The form of the other two differential boundary conditions 
may be obtained from the total free energy 
\begin{equation}\label{total}
F_{tot}[\phi]=F[\phi]+\int ds\gamma_{wf}(\phi),
\end{equation}
plus our knowledge of GNBC.
Here $\gamma_{wf}(\phi)$ is the interfacial free energy
per unit area at the fluid-solid boundary. We use
$\gamma_{wf}(\phi)=(\Delta\gamma_{wf}/2)\sin(\pi\phi/2)$
to denote
a smooth interpolation between $\pm\Delta\gamma_{wf}/2$,
with
$\Delta\gamma_{wf}=\gamma_{wf}(\phi_+)-\gamma_{wf}(\phi_-)$ 
given by $-\gamma\cos\theta_s^{surf}$ (Young's equation).
It should be noted that the form of the smooth interpolation
has very little effect on the final results. Hence we have chosen
a simple interpolation function.
Similar to Eq. (\ref{ch-force}), the change in $F_{tot}$ due to the
displacement of the molecules from $\bf r$  to 
${\bf r}'={\bf r}+{\bf u}({\bf r})$ is given by
\begin{equation}\label{total-force}
\delta F_{tot}=-\int d{\bf r}\left[{\bf g}\cdot{\bf u}\right] +\int
ds\left[\tilde{\sigma}_{ni}^Yu_i\right], 
\end{equation}
where
\begin{equation}\label{un-young-stress} 
\tilde{\sigma}_{ni}^Y=
-\left[ K\partial_n\phi+\displaystyle\frac
{\partial\gamma_{wf}(\phi)}{\partial\phi} \right]\partial_i\phi, 
\end{equation}
is the uncompensated Young stress \cite{jacqmin} (see below).
The continuum (differential) form of the GNBC (\ref{gnbc3}) 
is therefore given by
\begin{equation}\label{he3}
\beta v_x^{slip}=-\tilde{\sigma}_{nx}(0)=-\eta\left[\partial_n
v_x\right](0) +\left[ L(\phi)\partial_x\phi \right](0),
\end{equation}
where $\left[ L(\phi)\partial_x\phi \right](0)$, with
$L(\phi)=K\partial_n\phi+\partial\gamma_{wf}(\phi)/\partial\phi$,
is the differential expression for
$-\tilde{\sigma}_{nx}^Y(0)=-{\sigma}_{nx}^Y(0)+{\sigma}_{nx}^0(0)$
in equation (\ref{gnbc3}).
Here $K\partial_n\phi\partial_x\phi$ is $-{\sigma}_{nx}^Y(0)$
as seen in Eq. (\ref{young-stress}),
and $[\partial\gamma_{wf}(\phi)/\partial\phi]\partial_x\phi
=\partial_x\gamma_{wf}(\phi)$ \cite{landau} equals to 
${\sigma}_{nx}^0(0)$, in accordance with the static force balance 
relation $\partial_x\gamma_{wf}(\phi)-{\sigma}_{nx}^0(0)=0$.
From $\int_{\rm int}dx[K\partial_n\phi\partial_x\phi](0)
=\gamma\cos\theta_d^{surf}$ \cite{surface-angle} and
$\int_{\rm int}dx\partial_x\gamma_{wf}=-\gamma\cos\theta_s^{surf}$,
we see that 
$$\int_{\rm int}dx\left[ L(\phi)\partial_x\phi \right](0)
=\gamma(\cos\theta_d^{surf}-\cos\theta_s^{surf}),$$ 
in agreement with $\left[ L(\phi)\partial_x\phi \right](0)$
being the uncompensated Young stress.

Another boundary condition may be inferred from the fact
that $L(\phi)=0$ is the Euler-Lagrange equation at the fluid-solid
boundary for minimizing the total free energy $F_{tot}[\phi]$.
That is, $L(\phi)=0$ corresponds with
the equilibrium (static) condition where
$\partial\phi/\partial t+{\bf v}\cdot\nabla\phi=0$.
The boundary relaxation dynamics of $\phi$ is plausibly assumed to be
the first-order extension of that correspondence for a nonzero $L(\phi)$:
\begin{equation}\label{he4}
{\partial\phi\over\partial t}+{\bf v}\cdot\nabla\phi=
-\Gamma \left[ L(\phi)\right],
\end{equation}
where
$\Gamma$ is a (positive) phenomenological parameter.

\section{Comparison of MD and continuum hydrodynamics results}
\label{comparison}
\begin{figure}
\centerline{\psfig{figure=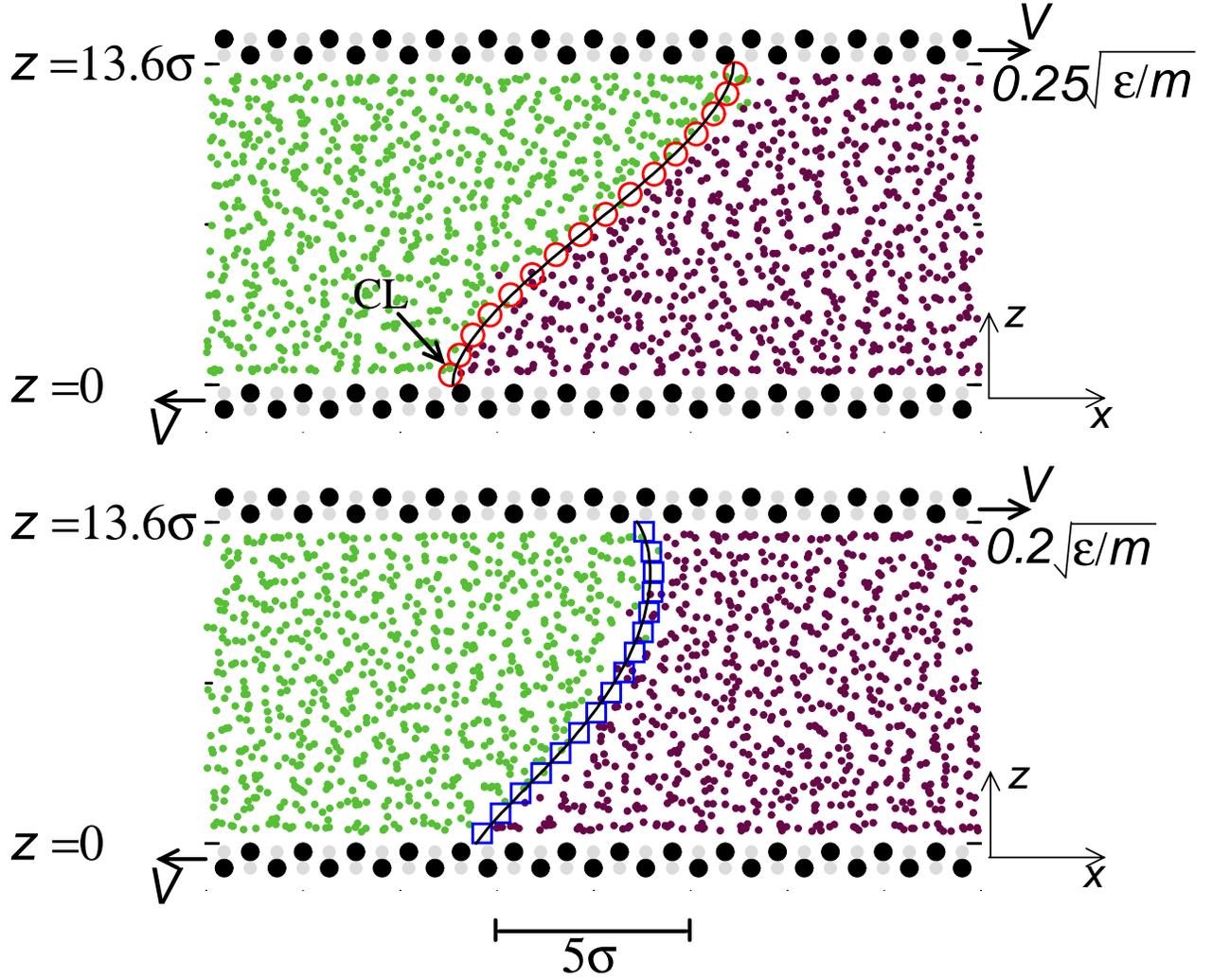,height=14.0cm}}
\bigskip
\caption{Segments of the MD simulation sample for
the immiscible Couette flows. The colored dots indicate
the instantaneous molecular positions of the two fluids
projected onto the $xz$ plane. The black/gray circles
denote the wall molecules. The upper panel illustrates
the symmetric case; the lower panel illustrates the asymmetric case.
The red circles and the blue squares represent the time-averaged
interface profiles, defined by $\rho_1=\rho_2$ ($\phi=0$),
for the two cases. The black solid lines are the
interface profiles calculated from the continuum hydrodynamic model
with the GNBC.}\label{fig1}
\end{figure}

\begin{figure}
\centerline{\psfig{figure=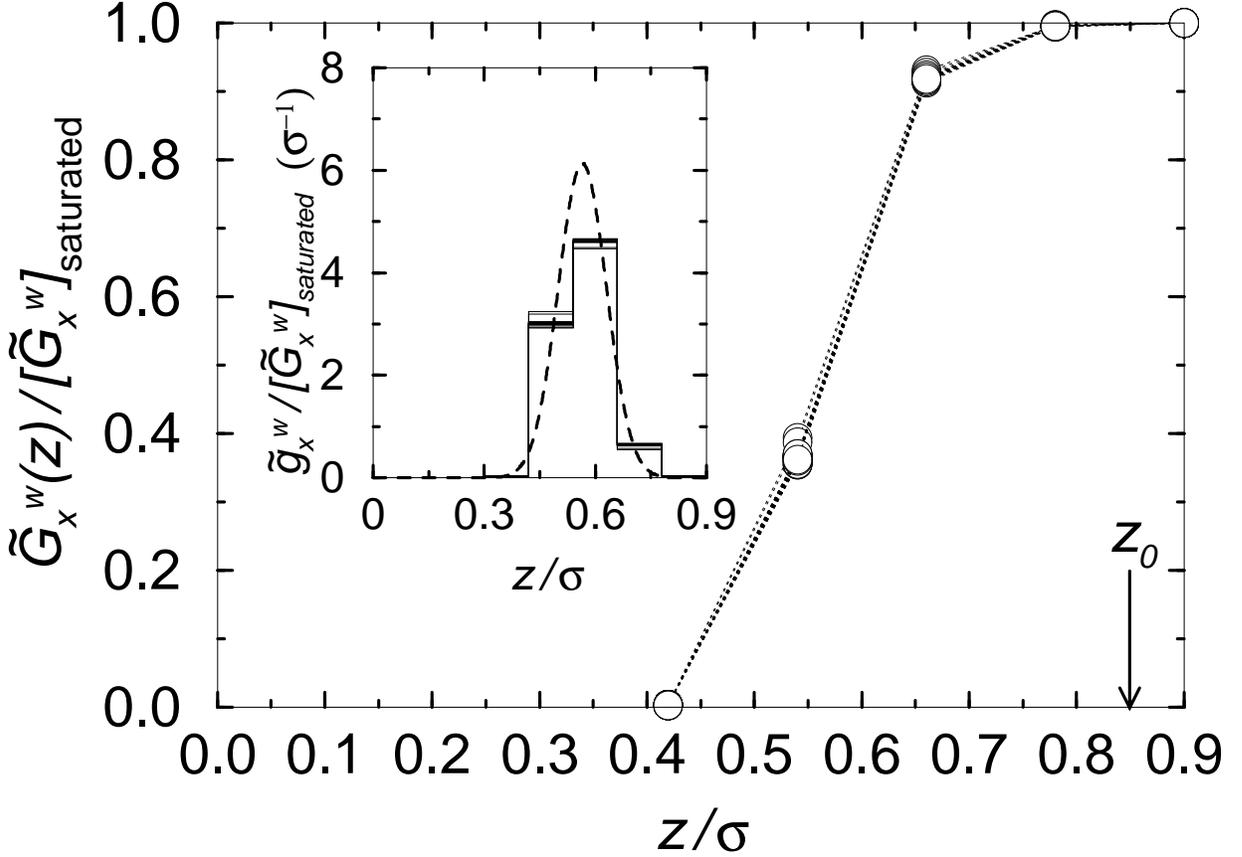,height=12.0cm}}
\bigskip
\caption{By subdividing the boundary layer into thin sections,
we plot the accumulated wall force per unit wall area
as a function of distance $z$ away from the boundary.
Here $\tilde{G}_x^w(z)$ is defined by 
$\tilde{G}_x^w(z)=\int_0^zdz'\tilde{g}_x^w(z')$
where $\tilde{g}_x^w$ is the density of tangential wall force.
For different $x$ positions, the absolute value of
the saturating total wall force is different.
However, when normalized by the corresponding
saturated total wall force per unit area at each $x$,
all points fall on a universal curve, nearly independent of $x$.
It is seen that at $z=z_0$
the wall force has reached its saturation value.
Inset: Tangential wall force density plotted as a function of 
distance $z$ away from the boundary. The solid lines are
averaged $\tilde{g}_x^w$ in thin sections at different $x$, 
normalized by the corresponding saturated total wall force 
per unit area. The dashed line is a smooth Gaussian fit.
In the sharp boundary limit this peaked wall force density 
is approximated by $\tilde{G}_x^w\delta(z)$.}\label{fig2}
\end{figure}

\begin{figure}
\centerline{\psfig{figure=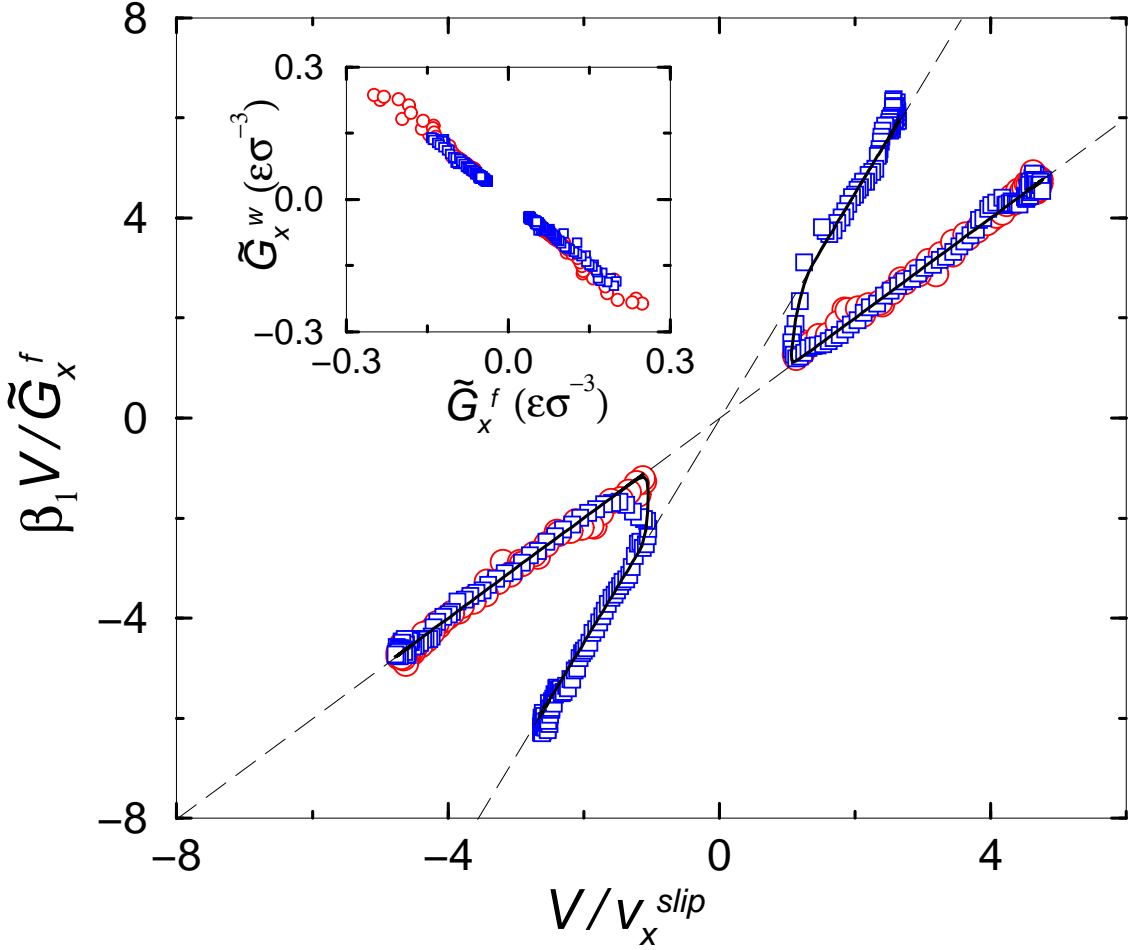,height=13.0cm}}
\bigskip
\caption{$\beta_1 V/\tilde{G}_x^f$ plotted as a function of
$V/v_x^{slip}$. Symbols are MD data measured in the BL at
different $x$ locations, where the red circles denote the symmetric
case and the blue squares the asymmetric case.
The solid lines were calculated from Eq. (\ref{gnbc1}) with values
of $\beta_{1,2}$ and the expression of $\beta$ given in the text.
The statistical errors of the MD data are about the size of the symbols.
The upper-right data segment corresponds to the lower boundary,
whereas the lower-left segment corresponds to the upper boundary.
The slopes of the two dashed lines are given by $\beta_{1,2}^{-1}$.
Inset: $\tilde{G}_x^w$ plotted as a function of $\tilde{G}_x^w$,
measured in the two BL's at different values of $x$.
The symbols have the same correspondence as in the main figure.
The data are seen to lie on a straight line with a slope of
$-1$, indicating $\tilde{G}_x^w+\tilde{G}_x^f=0$.}\label{fig3}
\end{figure}

\begin{figure}
\centerline{\psfig{figure=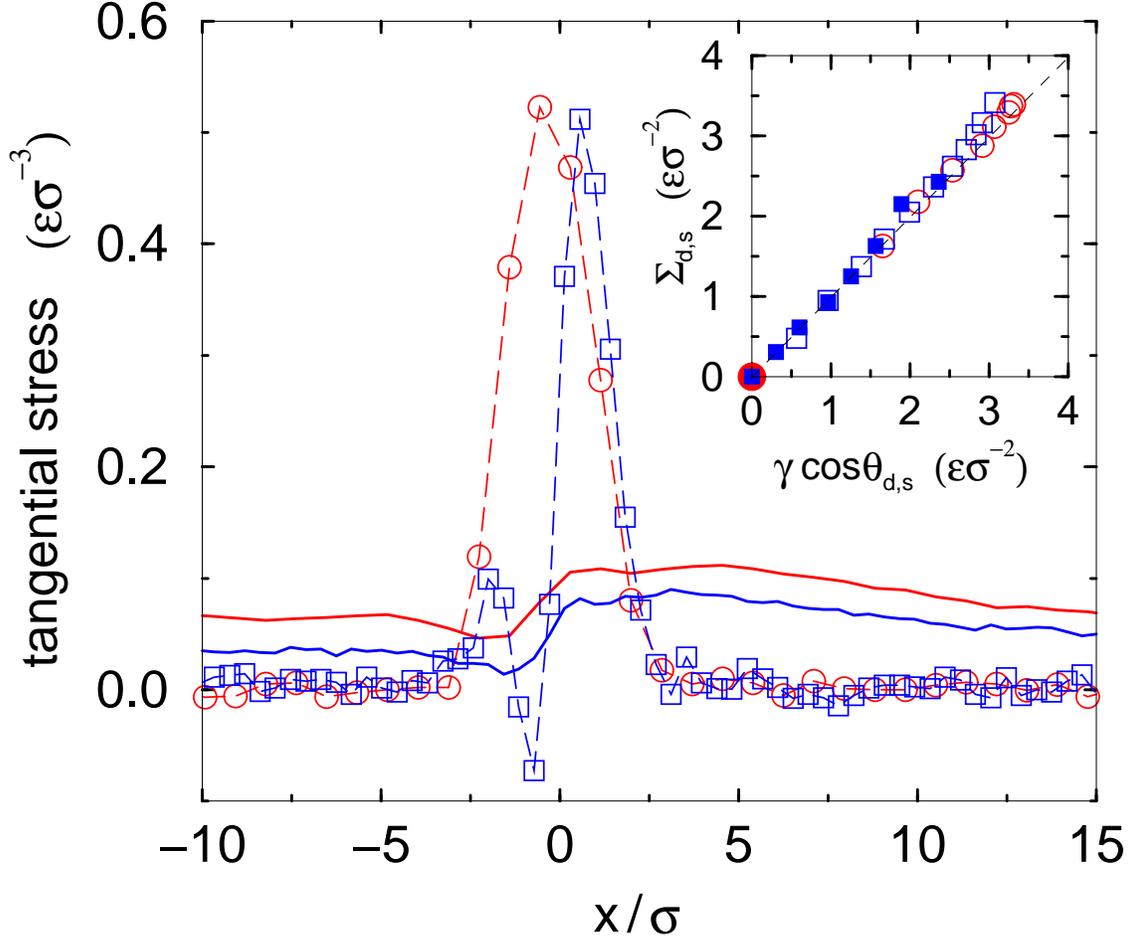,height=13.0cm}}
\bigskip
\caption{Two components of the dynamic tangential stress at $z=z_0$,
plotted as a function of $x$.
The dashed lines denote $\tilde{\sigma}_{zx}^Y$;
solid lines represent the viscous component.
Here red indicates the symmetric case and blue the asymmetric case.
In the CL region the non-viscous component is one order of magnitude larger
than the viscous component.  The difference between the two components,
however, diminishes towards the boundary, $z=0$, due to the large
interfacial pressure drop (implying a large curvature) in the BL,
thereby pulling $\theta_d$ closer to $\theta_s$.  Inset:
$\Sigma_{d,s}$ plotted as a function of $\gamma\cos\theta_{d,s}$
at different values of $z$.
Here $\Sigma_d=-\int dx(\sigma_{nx}-\sigma_{nx}^v)$,
$\Sigma_s=-\int dx\sigma_{nx}^0$, and $\theta_{d,s}$ was measured
from the time-averaged interfacial profiles (Fig. \ref{fig1}).
The red circles denote the symmetric case,
the blue squares the asymmetric case,
the solid blue squares the asymmetric static case,
and the single solid red circle at the origin denotes
the symmetric static case. The data are seen to follow a straight
(dashed) line with slope $1$,
indicating $\Sigma_{d,s}=\gamma\cos\theta_{d,s}$.}\label{fig4}
\end{figure}

\begin{figure}
\centerline{\psfig{figure=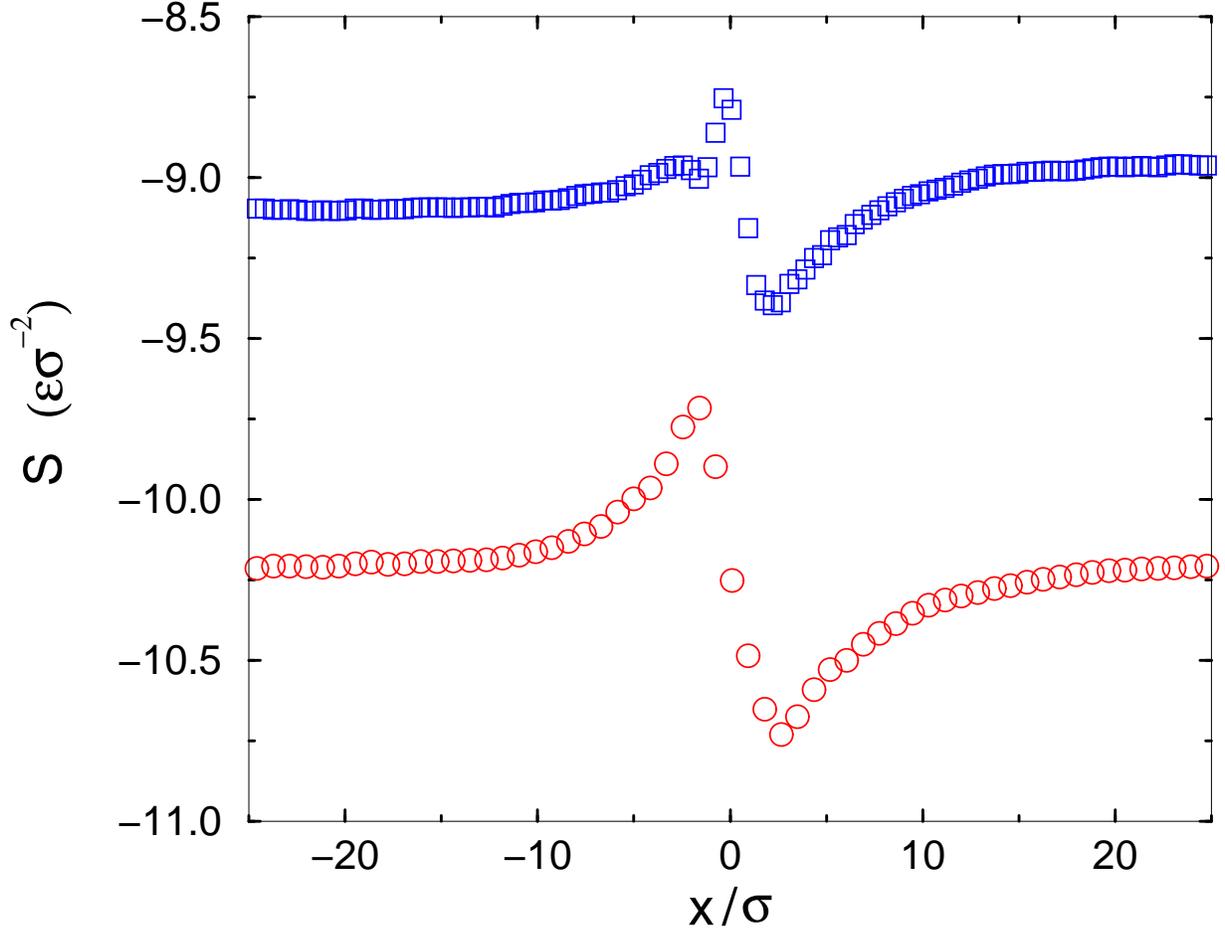,height=13.0cm}}
\bigskip
\caption{$S=\int_0^{z_0}\tilde{\sigma}_{xx}(z)dz=
\int_0^{z_0}\left[\sigma_{xx}(z)-\sigma_{xx}^0(z)\right]dz$ plotted
as a function of $x$. Here red circles denote the symmetric case
and blue squares the asymmetric case. 
For clarity, $\sigma_{xx}^0$ was vertically displaced such that
$\sigma_{xx}^0=0$ far from the interface in the symmetric case, and 
for the asymmetric case, $\sigma_{xx}^0=0$ at the center of the interface.
}\label{fig5}
\end{figure}

\begin{figure}
\centerline{\psfig{figure=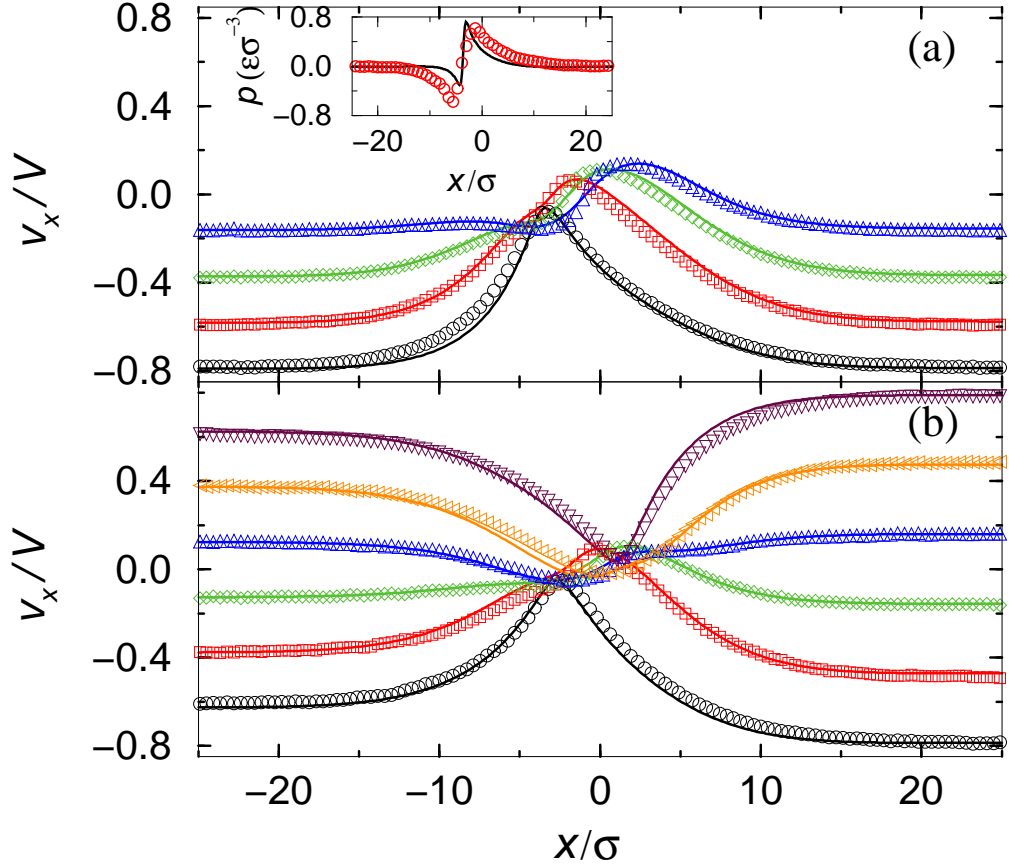,height=12.0cm}}
\bigskip
\caption{Comparisons between the MD (symbols) and the continuum
hydrodynamics (solid lines) results for the Couette flow,
the latter calculated with the GNBC and values of
$M=0.023\sigma^4/\sqrt{m\epsilon}$ and
$\Gamma=0.66\sigma/\sqrt{m\epsilon}$.
(a) The $v_x$ profiles for the symmetric case
($V=0.25(\epsilon/m)^{1/2}$ and $H=13.6\sigma$) at different $z$ planes.
The profiles are symmetric about the center plane,
hence only the lower half is shown for
$z=0.425\sigma$ (black circles),
$2.125\sigma$ (red squares),
$3.825\sigma$ (green diamonds), and
$5.525\sigma$ (blue triangles).
(b) The $v_x$ profiles for the asymmetric case
($V=0.2(\epsilon/m)^{1/2}$ and $H=13.6\sigma$)
at $z=0.425\sigma$ (black circles), $2.975\sigma$ (red squares),
$5.525\sigma$ (green diamonds), $8.075\sigma$ (blue triangles),
$10.625\sigma$ (yellow triangles), $13.175\sigma$ (maroon triangles).
For the boundary layers, $v_x=0$ means complete slip.
Inset: Pressure variation in the BL for the symmetric case.
The solid line represents the BL-averaged hydrodynamic pressure
$z_0^{-1}\int_0^{z_0}p(z)dz$ from the continuum model,
and red circles denote $z_0^{-1}\int_0^{z_0}\tilde{\sigma}_{xx}(z)dz$
measured in MD simulations (see Fig. \ref{fig5}).
Note the fast variation across the interface.
The interfacial pressure drop in the BL is a factor $5-10$ larger than
that in the middle of the sample, implying large curvature.}\label{fig6}
\end{figure}

Motivated by the methods presented in \cite{liu,weinan},
a second order scheme is designed to solve the CH hydrodynamic model,
comprising the dynamic equations and the four boundary conditions. Details
of the numerical algorithm are presented in Appendix \ref{app-numerical}.
Besides those parameters which can be directly obtained from MD simulations,
$M$ and $\Gamma$ are treated as fitting parameters, determined by 
comparison with MD results (values given in the caption to Fig. \ref{fig6}).
In Figs. \ref{fig1} and \ref{fig6} we show
that the continuum model can indeed quantitatively reproduce the interface and
velocity profiles from MD simulations, including the near-complete slip
($v_x\approx 0$ in Fig. \ref{fig6}) of the CL, the fine features 
in the molecular-scale vicinity of the CL, 
and the fast pressure variation in the BL (inset to Fig. \ref{fig6}), 
with its implied large interfacial curvature.
We wish to emphasize that for the comparison with the symmetric case,
the parameters in the continuum model, including those in the GNBC,
are directly obtained from the MD simulations, whose velocity profiles are then
fitted by those from the hydrodynamic calculations with optimized $M$
and $\Gamma$ values. Thus the comparison with the asymmetric (Couette) case,
with $\beta_2$ directly evaluated from MD simulation data, is 
{\it without adjustable parameters}. 
We have also obtained $\theta_d^{surf}=88.1^\circ$ and $62.8^\circ$
for the symmetric and asymmetric (the lower boundary) cases
shown in Figs. \ref{fig1}a and \ref{fig1}b, respectively. 
Both are in excellent agreement with their extrapolated values in MD simulations.
For the upper boundary in the asymmetric case, 
our calculated $\theta_d^{surf}=65.2^\circ$, which differs somewhat from the MD 
extrapolation value of $64.5\pm 0.5^\circ$. This difference is a reflection
of the discrepancy seen in Fig. \ref{fig3}. However, it is noteworthy that
the difference in the dynamic contact angles does not show up in the
velocity profiles, which agree well.

\begin{figure}
\centerline{\psfig{figure=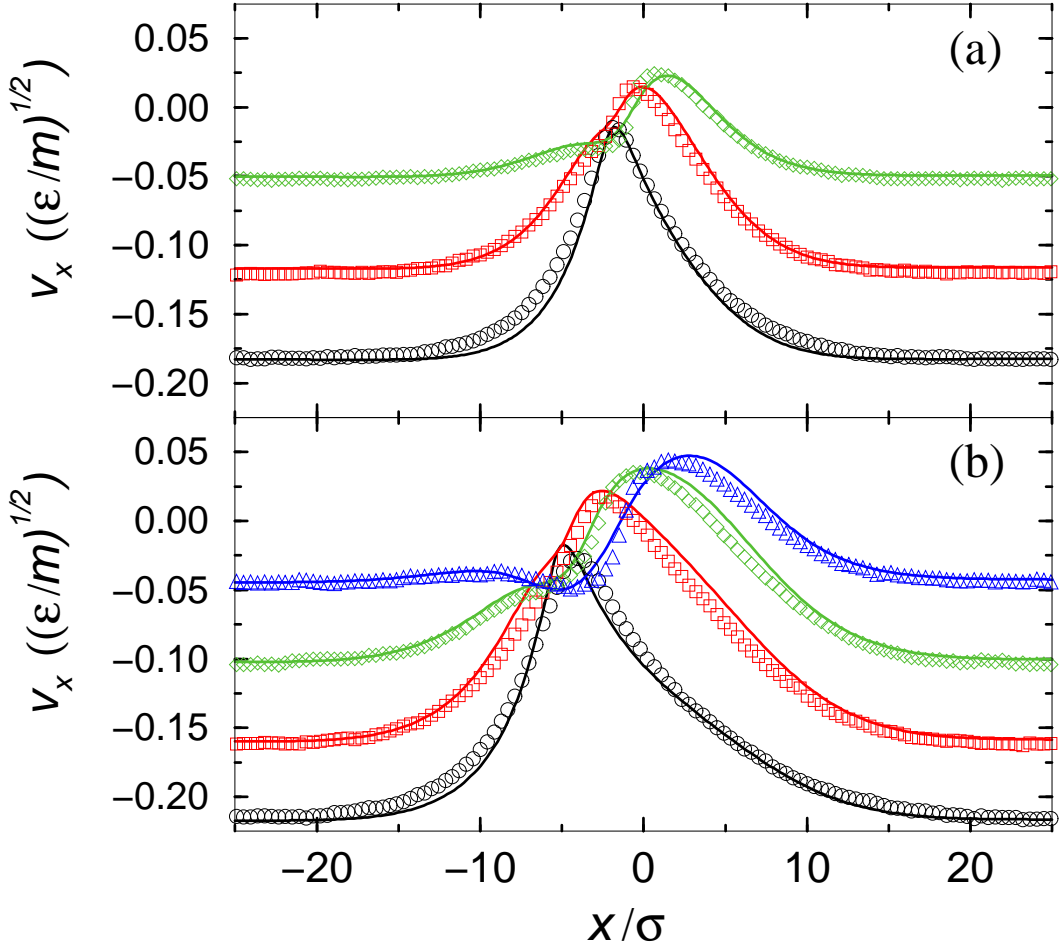,height=13.0cm}}
\bigskip
\caption{Comparisons between the MD (symbols) and the continuum
hydrodynamics (solid lines) results for two symmetric cases
in the Couette flow geometry. Compared with Fig. \ref{fig6}a,
$V$ and $H$ have been varied, respectively, but the continuum
results are calculated with the same set of parameters and the GNBC.
The profiles are symmetric about the center plane,
hence only the lower half is shown. (a) The $v_x$ profiles
for $V=0.25(\epsilon/m)^{1/2}$ and $H=10.2\sigma$,
shown at $z=0.425\sigma$ (black circles), $2.125\sigma$ (red squares),
and $3.825\sigma$ (green diamonds). (b) The $v_x$ profiles
for $V=0.275(\epsilon/m)^{1/2}$ and $H=13.6\sigma$, shown at
$z=0.425\sigma$ (black circles), $2.125\sigma$ (red squares),
$3.825\sigma$ (green diamonds), and
$5.525\sigma$ (blue triangles).
}\label{fig7}
\end{figure}

\begin{figure}
\centerline{\psfig{figure=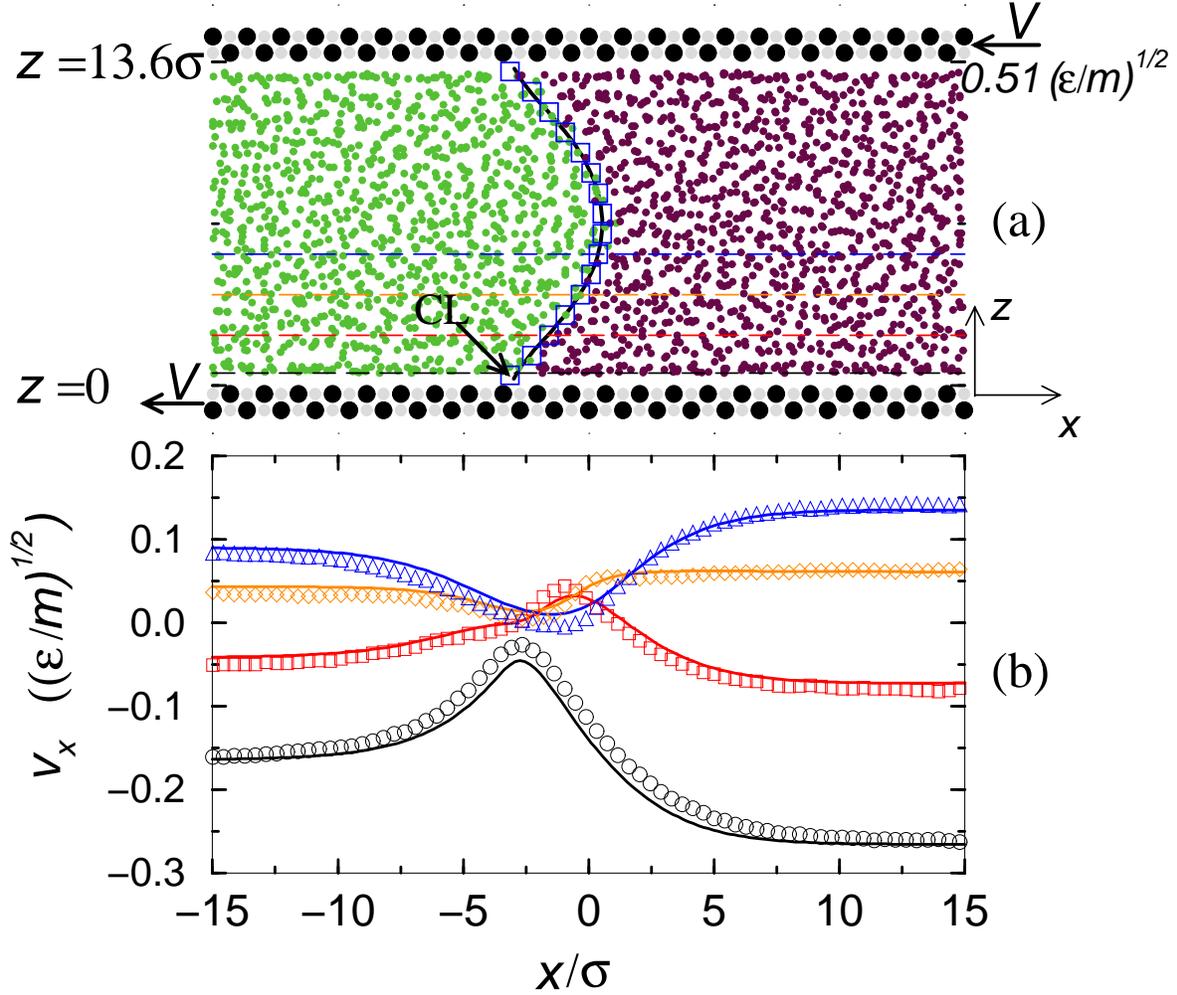,height=14.0cm}}
\bigskip
\caption{Comparisons between the MD (symbols) and the continuum
hydrodynamics (solid lines) results for an asymmetric case
in the Poiseuille flow geometry. Compared with Fig. \ref{fig6}b,
the type of flow has been changed, but the continuum results are
calculated with the same set of parameters. (a) A segment of the
instantaneous configuration in the MD simulation. The two walls, separated
by $H=13.6\sigma$, move at a constant speed $V=0.51(\epsilon/m)^{1/2}$
in the $-x$ direction in order to maintain a time-independent steady-state
interface, with $mg_{ext}=0.05\epsilon/\sigma$ applied in the $x$ direction.
The symbols have the same correspondence as those in Fig. \ref{fig1}b.
The black solid line is the interface profiles calculated
from the continuum hydrodynamic model. The colored dashed lines
indicate the $z$ coordinates of the $v_x$ profiles shown in (b).
(b) The $v_x$ profiles at $z=0.425\sigma$ (black circles),
$2.125\sigma$ (red squares), $3.825\sigma$ (green diamonds),
and
$5.525\sigma$ (blue triangles). The profiles are symmetric about
the center plane, hence only the lower half is shown.}\label{fig8}
\end{figure}

To further verify that the boundary conditions and the parameter values
are local properties and hence applicable to flows with 
different macroscopic conditions, we have varied the wall speed $V$, 
the system size $H$, and the flow geometry to check that 
the same set of parameters plus the GNBC are valid for reproducing 
(a) the velocity profiles from a different set of Couette-flow simulations 
in the symmetric configurations, shown in Fig. \ref{fig7}, 
as well as (b) the velocity profiles of the Poiseuille flow simulations 
in the asymmetric case, shown in Fig. \ref{fig8}. The remarkable overall 
agreement in all cases affirms the validity of GNBC and 
the hydrodynamic model \cite{bc}, as well as justifies the replacement of the 
diffuse fluid-solid boundary (force density) by a sharp boundary.

\begin{figure}
\bigskip
\centerline{\psfig{figure=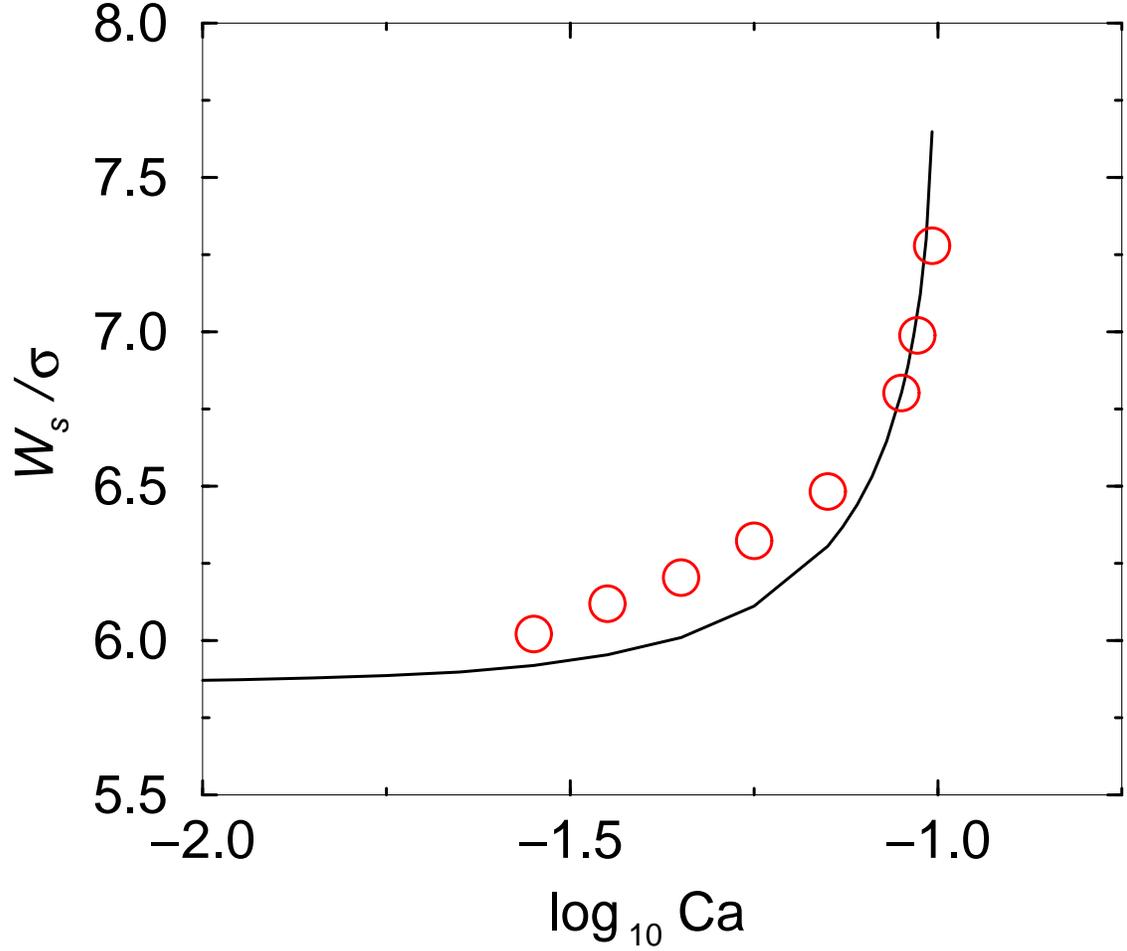,height=13.0cm}}
\caption{Width for the moving CL region, $W_s$, plotted
as a function of the capillary number $Ca=\eta V/\gamma$ 
for the symmetric case by varying $V$ and keeping $H=13.6\sigma$.
We note that for most of the MD data measured in the symmetric case,
$Ca\simeq 0.088$. Solid line was calculated from the 
immiscible hydrodynamic model employing the GNBC;
red circles denote the MD results.}\label{fig9}
\end{figure}

Another comparison is the dissipation incurred by the moving CL
in the Couette flow geometry. Rate of heat generation per unit wall area
is given by $|\tilde{G}_x^w|V=\beta |v_x^{slip}|V$.
From this we have to subtract a small but constant relative
slipping away from the interface, $v_0^{slip}=2Vl_s/(H+2l_s)$, where
$l_s=\eta/\beta$ is a slip length for fluid 1 if $\phi<0$ and 
for fluid 2 if $\phi>0$. The resulting heat generation rate due
to the CL is $\beta V^2 W_s L$ (for one wall), where $L$ 
is the length of the CL and $W_s$ defines the width of the CL region:
\begin{equation}\label{width}
W_s=\displaystyle\frac{1}{V}\int\left( |v_x^{slip}|-v_0^{slip} \right)dx.
\end{equation}
Thus CL dissipation is equivalent to a segment, $\sim H(W_s/l_s)$, 
of dissipation by single phase flow. Figure \ref{fig9} shows the variation of
$W_s$ as a function of capillary number $Ca=\eta V/\gamma$ 
for the symmetric case of Couette flow.  Close to $Ca\simeq 0.1$ 
the value of $W_s$ increases rapidly, in good agreement with the MD results,
and beyond which the continuum model failed to converge. This corresponds to
the breakup of the interface observed in MD simulations \cite{breakup}.

\section{Concluding remarks}
\label{remark}
In summary, we have found for the first time the boundary condition that yields
near-complete slipping of the CL, in good agreement with MD results on the
molecular scale. 
It should also be noted, however, that the present continuum formulation can not
calculate fluctuation effects that are important in MD simulations. Long range
interactions, e.g., that due to van der Waals interaction, have also been ignored.
The latter is potentially important in the calculations involving wetting layers.

\section*{acknowledgments}
Partial support from HKUST's EHIA funding and Hong Kong RGC
grants HKUST 6176/99P and 6143/01P is hereby acknowledged.

\appendix
\section{wall-fluid interactions}\label{app-wall-fluid}
We have measured both the tangential and normal components of the wall force
exerted on the fluids. Both components vary along the $z$ direction 
and saturate somewhere away from the fluid-solid boundary. 
The tangential component saturates (by $99.8\%$) at $z=z_0$, 
which is well inside the wall-fluid interaction range 
($z_0=0.85\sigma$, smaller than the cut-off distance $r_c=2.5\sigma$ 
for the wall-fluid interaction potential $U_{wf}$).
On the other hand, the normal component is $87\%$ of the saturation value
at $z=z_0$ and $99.8\%$ at $z=2z_0$. The different saturation
ranges of the tangential and normal components may be understood as follows.

For a fluid molecule close to the solid wall,
the interaction with one particular (the closest) wall molecule 
can be much stronger than all the others. As this fluid molecule moves
laterally but still remaining its close proximity to the wall, it would thus
experience a strong periodic modulation in its interaction with the wall.
This lateral inhomogeneity is an important source for the tangential 
component of the wall force.
Away from the fluid-solid boundary, each fluid molecule can interact with 
many wall molecules on a nearly equal basis. 
Thus the modulation amplitude of the wall potential would clearly decrease with 
increasing distance from the wall. Hence the tangential wall force tends to
saturate at the relatively short range of $z\simeq z_0$. 
On the contrary, the normal wall force directly arises from 
the wall-fluid interaction, independent of whether the wall potential is 
``rough'' or not. Consequently, the normal wall force saturates much slower 
than the tangential component.

The MD results presented in this paper were obtained from simulations
using solid walls constructed by two [001] planes of an fcc lattice.
We have also carried out MD simulations using thicker confining walls.
First we changed the number of molecular layers ([001] planes of fcc lattice)
from two to four in constructing each of the two walls. 
The wall-fluid interaction potential $U_{wf}$ were still cut off
at $r_c=2.5\sigma$. It turned out that neither component of the wall force 
shows any noticeable change. The reason is that 
for the tangential component, the two outer planes 
are too distant to contribute to the roughness of the wall potential, while
for the normal component, the fluid molecules closest to the wall are 
separated from the two outer planes by a distance $\ge r_c$. 
Consequently, both the interface and velocity profiles do not show
any noticeable change.

Additional wall layers do not contribute to the perceived modulation of 
the wall potential by the fluid molecules. Nevertheless, they can still affect 
the tangential wall force by modifying the organization of the fluid molecules
near the wall. Such organization is governed by the wall-fluid interaction 
and can be greatly influenced by the normal wall force.
To see the effects of normal wall force due to additional wall layers, 
we used four [001] planes of an fcc lattice plus a half-space continuum
in constructing a wall. The first four solid layers show the atomic structure
detectable by the fluid molecules while the half-space continuum models the
deeper solid layers. The wall-fluid interaction was modelled as follows.
For an in-range pair of fluid and wall molecules separated by a distance 
$r<r_c$, the interaction potential is still $U_{wf}$.
Here the wall molecule must be from one of the four solid layers.
In addition to this short-range interaction, the fluid molecules
can also experience the long-range interaction potential due to (1) the
distant wall molecules in the four solid layers and (2) the continuum.
For (1) we integrated the $1/r^6$ term in $U_{wf}$ over the
out-of-range ($r>r_c$) area of the solid layers while for (2) we integrated
the same term over the half-space continuum. 
According to this model, only the in-range ($r<r_c$) part of 
the solid wall shows atomic structure to a fluid molecule while the
out-of-range ($r>r_c$) part is effectively a half-space continuum.

We found that the effect of the long-range normal wall force (for $\delta_{wf}>0$)
is to attract the fluid molecules to the wall. In fact the average number 
density in the BL can increase by $3-4\%$ once the long-range force is included.
As a result, the slip coefficient $\beta_{1(2)}$ increases by $\sim 5-15\%$.
This results in small but visible changes in the interface and velocity profiles.

These tests have convinced us that by using two [001] planes of an fcc lattice
to model the solid wall, we have captured the dominant wall-fluid interaction.
In fact, using two molecular layers to model the solid wall has been extensively
practiced in the past MD simulations \cite{koplik,robbins,nbc,bc,two-layers},
although in some instances more molecular layers have also been used 
\cite{more-layers}, where the accurate modeling of the normal component
of the wall-fluid interaction force is important.

\section{stress measurements in MD simulations}\label{app-md-stress}
\subsection{Microscopic formula of Irving and Kirkwood}
Irving and Kirkwood \cite{stress} have shown that in the hydrodynamic equation
of motion (momentum transport), the stress tensor (flux of momentum) may be
expressed in terms of molecular variables as
\begin{equation}\label{total-stress}
{\mbox{\boldmath$\sigma$}}({\bf r},t)=
{\mbox{\boldmath$\sigma$}}_K({\bf r},t)+
{\mbox{\boldmath$\sigma$}}_U({\bf r},t),
\end{equation}
where ${\mbox{\boldmath$\sigma$}}_K$ is the kinetic contribution to 
the stress tensor, given by
\begin{equation}\label{K-stress}
{\mbox{\boldmath$\sigma$}}_K({\bf r},t)=-\left\langle\sum_i m_i 
\left[\displaystyle\frac{{\bf p}_i}{m_i}-{\bf V}({\bf r},t)\right]
\left[\displaystyle\frac{{\bf p}_i}{m_i}-{\bf V}({\bf r},t)\right]
\delta({\bf x}_i-{\bf r}),f\right\rangle,
\end{equation}
and ${\mbox{\boldmath$\sigma$}}_U$ is the contribution of 
intermolecular forces to the stress tensor, given by
\begin{equation}\label{U-stress}
{\mbox{\boldmath$\sigma$}}_U({\bf r},t)=-\displaystyle\frac{1}{2}
\left\langle\sum_i\sum_{j\ne i}({\bf x}_i-{\bf x}_j){\bf F}_{ij}
\delta({\bf x}_i-{\bf r}),f\right\rangle.
\end{equation}
Here $m_i$, ${\bf p}_i$, and ${\bf x}_i$ are respectively
the mass, momentum, and position of molecule $i$,
${\bf V}({\bf r},t)$ is the local average velocity, ${\bf F}_{ij}$
is the force on molecule $i$ due to molecule $j$,
$f$ is the probability distribution function 
$$f({\bf x}_1,\cdots,{\bf x}_N,
{\bf p}_1,\cdots,{\bf p}_N,t),$$ 
which satisfies the normalization condition
$$
\int{\rm d}{\bf x}_1\cdots{\rm d}{\bf x}_N
{\rm d}{\bf p}_1\cdots{\rm d}{\bf p}_Nf=1,$$
and the Liouville equation
$$
\displaystyle\frac{\partial f}{\partial t}=
-\sum_i\left[\displaystyle\frac{{\bf p}_i}{m_i}
\cdot\displaystyle\frac{\partial f}{\partial{\bf x}_i}
-{\mbox{\boldmath$\nabla$}}_{{\bf x}_i}U\cdot\displaystyle\frac
{\partial f}{\partial{\bf p}_i}\right],
$$
with $U$ being the potential energy of the system,
and 
$\langle \cdot\cdot\cdot,f \rangle$ means taking the average  
for a probability distribution function $f$.

Although widely employed in the stress measurements in MD simulations,
the above expression for ${\mbox{\boldmath$\sigma$}}_U$ 
(Eq. (\ref{U-stress})) represents only the leading term 
in an asymptotic expansion, accurate when the interaction range 
is small compared to the range of hydrodynamic variation
\cite{stress}. This can be seen as follows.

Consider that all the molecules interact via a pair potential
$U_{\rm pair}(R)$ such that the intermolecular force 
${\bf F}_{ij}=({\bf R}/R)U_{\rm pair}'(R)$ for
${\bf x}_j={\bf x}_i+{\bf R}$.
Accordingly, Eq. (\ref{U-stress}) can be rewritten as
\begin{equation}\label{U-stress-1}
{\mbox{\boldmath$\sigma$}}_U({\bf r},t) 
=\displaystyle\frac{1}{2}\int d{\bf R}
\displaystyle\frac{{\bf R}{\bf R}}{R}U_{\rm pair}'(R)
\rho^{(2)}({\bf r},{\bf r}+{\bf R},t),
\end{equation}
where $\rho^{(2)}$ is the pair density defined by
$$
\rho^{(2)}({\bf r}_1,{\bf r}_2,t)=
\sum_{i\ne j}\left\langle \delta({\bf r}_i-{\bf r}_1)
\delta({\bf r}_j-{\bf r}_2),f \right\rangle.
$$
It has been shown (see the appendix in Ref. \cite{stress}) 
that according to the definition that
$d{\bf S}\cdot{\mbox{\boldmath$\sigma$}}_U$ is 
the force acting across $dS$,
the full expression for ${\mbox{\boldmath$\sigma$}}_U$
is given by
\begin{equation}\label{U-stress-2}
{\mbox{\boldmath$\sigma$}}_U({\bf r},t)=
\displaystyle\frac{1}{2}\int d{\bf R}
\displaystyle\frac{{\bf R}{\bf R}}{R}U_{\rm pair}'(R)
\left[\int_0^1 d\alpha
\rho^{(2)}({\bf r}-\alpha{\bf R},{\bf r}-\alpha{\bf R}+{\bf R},t)
\right].
\end{equation}
It is readily seen that Eq. (\ref{U-stress-1}) may be obtained
from Eq. (\ref{U-stress-2}) by keeping only the lowest order term 
in a Taylor's series in $\alpha$, i.e.,
$$\rho^{(2)}({\bf r}-\alpha{\bf R},{\bf r}-\alpha{\bf R}+{\bf R},t)
\approx \rho^{(2)}({\bf r},{\bf r}+{\bf R},t).$$
That means ${\bf R}\cdot{\mbox{\boldmath$\nabla$}}_{\bf r}
\rho^{(2)}({\bf r},{\bf r}+{\bf R},t)$ must be negligible
compared with $\rho^{(2)}({\bf r},{\bf r}+{\bf R},t)$. Here
$\bf R$ is on the order of the range of intermolecular
force. This approximation, however, can not be justified
at the fluid-fluid or the wall-fluid interface, where
${\bf R}\cdot{\mbox{\boldmath$\nabla$}}_{\bf r}
\rho^{(2)}({\bf r},{\bf r}+{\bf R},t)$ can be comparable 
in magnitude to $\rho^{(2)}$.

\subsection{Stress measurement in the boundary layer}
\begin{figure}
\bigskip
\centerline{\psfig{figure=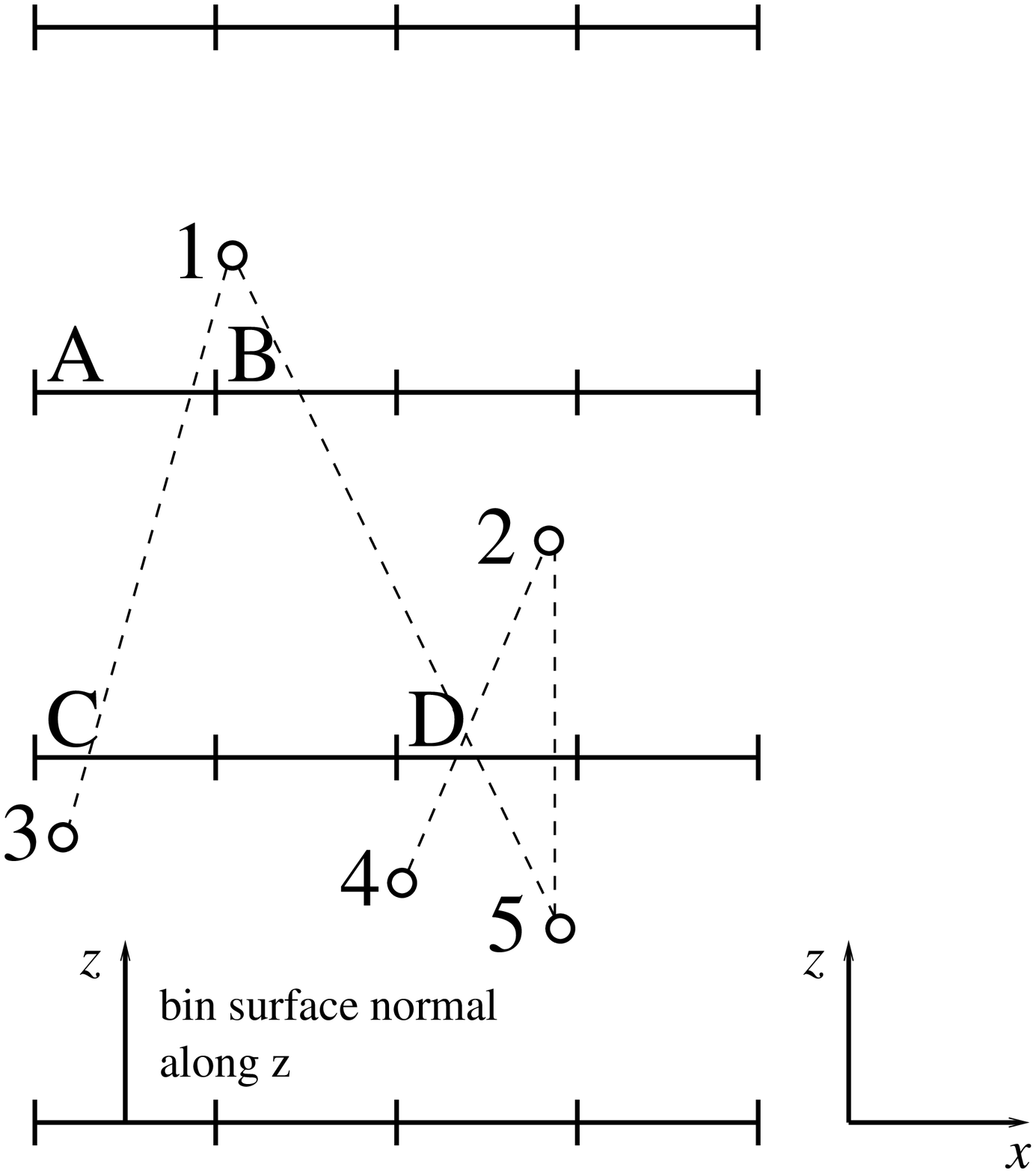,height=15.0cm}}
\caption{Schematic illustration of measuring the $zx$ component of
${\mbox{\boldmath$\sigma$}}_U$. The horizontal solid lines 
(separated by short vertical lines) represent bin surfaces 
with surface normals along the $z$ direction. 
Circles denote fluid molecules. The dashed lines 
connect pairs of interacting molecules. Here the bin surfaces and 
the molecules are projected onto the $xz$ plane. Molecules that 
appear to be close to each other may not be in the 
interaction range if their distance along $y$ is too large.
A pair of interacting molecules may act across more than one
bin surface. Here the (1,3) pair acts across the surfaces A and C
while the (1,5) pair acts across the surfaces B and D.
At each bin surface the stress measurement must run over all the
pairs that act across that surface. For surface D, there are 
three pairs of interacting molecules (1,5), (2,4), and (2,5)
that contribute to the $zx$ component of 
${\mbox{\boldmath$\sigma$}}_U$.}\label{fig-stress}
\end{figure}

In the study of moving CL, it is of great importance to 
obtain the correct information about stress distributions
at both the fluid-fluid and the wall-fluid interfaces.
Therefore, we have directly measured the $x$ component of 
fluid-fluid interaction forces acting across 
the $x(z)$ bin surfaces, in order to obtain the $xx(zx)$ component
of ${\mbox{\boldmath$\sigma$}}_U$. 
For example, in measuring $\sigma_{Uzx}$ at a given $z$-direction
bin surface, we recorded all the pairs of molecules interacting
across that surface. Here ``interacting across'' means that
the line connecting a pair of molecules intersects 
the bin surface. For those pairs, we then computed 
$\sigma_{Uzx}$ at the given bin surface using
$$
\sigma_{Uzx}=\displaystyle\frac{1}{\delta s_z}\sum_{(i,j)}F_{ijx},
$$
where $\delta s_z$ is the area of $z$-direction bin surface,
$(i,j)$ indicate all possible pairs of molecules interacting across 
the bin surface, with molecule $i$ being ``inside of $\hat{\bf z}\delta s_z$''
and molecule $j$ being ``outside of $\hat{\bf z}\delta s_z$''
(molecule $i$ is below molecule $j$), and $F_{ijx}$ is the $x$ component 
of the force on molecule $i$ due to molecule $j$.
A schematic illustration is shown in Fig. \ref{fig-stress}.

For comparison, we have measured the $xx$ and $zx$ 
components of ${\mbox{\boldmath$\sigma$}}_U$ using the discrete 
version of Irving-Kirkwood expression (\ref{U-stress}):
$$
{\mbox{\boldmath$\sigma$}}_U=
-\displaystyle\frac{1}{2\delta v}\left\langle
\sum_i\sum_{j\ne i}({\bf x}_i-{\bf x}_j){\bf F}_{ij}
\right\rangle,
$$
where $\delta v$ is the volume of sampling bin, 
$i$ runs over fluid molecules in the sampling bin,
$j$ runs over fluid molecules in interaction with
molecule $i$, and $\langle\cdots\rangle$ means taking the
time average. We found that far from the the fluid-fluid and 
the wall-fluid interfaces, the results based on the
Irving-Kirkwood expression agree well with those from
direct force measurement, whereas near the fluid-fluid or 
the wall-fluid interface, the two results show 
appreciable differences (up to $50\%$), especially
for the $zx$ component at the fluid-fluid interface.

\subsection{Relation of MD-measured stresses\\ to the continuum 
hydrodynamic stress components}
\label{md-continuum-correspondence}
We want to note the correspondence between the MD-measured stresses
and the continuum hydrodynamic stress components. This correspondence
is essential to obtaining the microscopic contact angle $\theta_d^{surf}$, 
defined in the continuum hydrodynamic model but not directly measurable 
in MD simulations.

The GNBC for the diffuse BL is given by
\begin{equation}\label{extrap-gnbc}
\tilde{G}_x^f=\beta v^{\rm slip}_x=
\displaystyle\frac{\partial}{\partial x}
\int_0^{z_0}dz\left[\sigma_{xx}(z)-\sigma_{xx}^0(z)\right]+
\left[\sigma_{zx}(z_0)-\sigma_{zx}^0(z_0)\right],
\end{equation}
which involves only MD measurable quantities. 
To obtain the contact angle $\theta_d^{surf}$ from MD results, 
we need to interpret the MD-measured quantities in terms of
the various continuum variables in the hydrodynamic model.
In so doing it is essential to note the following.

(1) $\sigma_{xx}$ can be decomposed into a molecular component
and a hydrodynamic component: $\sigma_{xx}=T_{xx}+\sigma_{xx}^{HD}$.
Meanwhile, $\sigma_{xx}^0$ can be composed into the same molecular component
and a hydrostatic component: $\sigma_{xx}^0=T_{xx}+\sigma_{xx}^{HS}$.
The molecular component $T_{xx}$ exists even if there is no
hydrodynamic fluid motion or fluid-fluid interfacial curvature.
In particular, $T_{xx}$ in the BL depends on the wall-fluid interactions.
The change of the BL-integrated $T_{xx}$ across the fluid-fluid interface 
equals the change in the wall-fluid interfacial free energy, 
i.e., $\int_{\rm int}dx\displaystyle\frac{\partial}
{\partial x}\left[\int_0^{z_0}dzT_{xx}(z)\right]=\Delta\gamma_{wf}
=\gamma_{wf}(\phi_+)-\gamma_{wf}(\phi_-)$.
On the other hand, the hydrodynamic component $\sigma_{xx}^{HD}$
in $\sigma_{xx}$ results from the hydrodynamic fluid motion 
and fluid-fluid interfacial curvature. In the static ($V=0$ or $g_{ext}=0$) 
configuration, $\sigma_{xx}^{HD}$ becomes the hydrostatic component 
$\sigma_{xx}^{HS}$ in $\sigma_{xx}^0$.

(2) $\sigma_{zx}(z_0)$ can be decomposed into a viscous component 
plus a Young component:
$\sigma_{zx}(z_0)=\sigma_{zx}^v(z_0)+\sigma_{zx}^Y(z_0)$
with $\sigma_{zx}^v=\eta(\partial_zv_x+\partial_xv_z)$
and $\int_{\rm int}dx\sigma_{zx}^Y(z_0)=\gamma\cos\theta_d(z_0)$.

(3) $\sigma_{zx}^0(z_0)$ is the static Young stress: i.e.,
$\int_{\rm int}dx\sigma_{zx}^0(z_0)=\gamma\cos\theta_s(z_0)$.

With the help of the above relations, integration of 
Eq. (\ref{extrap-gnbc}) across the fluid-fluid interface yields
\begin{equation}\label{extrap-gnbc-int1}
\int_{\rm int}dx\beta v^{\rm slip}_x=
\Delta\left[\int_0^{z_0}dz\sigma_{xx}^{HD}\right]
+\int_{\rm int}dx\sigma_{zx}^v(z_0)+\gamma\cos\theta_d(z_0)
-\Delta\left[\int_0^{z_0}dz\sigma_{xx}^{HS}\right]
-\gamma\cos\theta_s(z_0),
\end{equation}
where $\Delta\left[\int_0^{z_0}dz\sigma_{xx}^{HD(HS)}\right]$ 
is the change of the $z$-integrated $\sigma_{xx}^{HD(HS)}$ across 
the interface. According to the Laplace's equation, the change of 
the hydrostatic $z$-integrated normal stress is directly related to 
the static $z$-integrated curvature ${\cal K}_s$:
\begin{equation}\label{extrap-static}
-\Delta\int_0^{z_0}dz\sigma_{xx}^{HS}
=\gamma{\cal K}_s=\gamma\left[\cos\theta_s(z_0)-\cos\theta_s^{surf}\right].
\end{equation}
Note that ${\cal K}_s$ vanishes in the symmetric case.
Substituting Eq. (\ref{extrap-static}) into Eq. (\ref{extrap-gnbc-int1})
then yields
\begin{equation}\label{extrap-gnbc-int2}
\int_{\rm int}dx\beta v^{\rm slip}_x=
\Delta\int_0^{z_0}dz\sigma_{xx}^{HD}+
\int_{\rm int}dx\sigma_{zx}^v(z_0)+\gamma\cos\theta_d(z_0)
-\gamma\cos\theta_s^{surf}.
\end{equation}
If interpreted in the continuum hydrodynamic formulation with a
sharp fluid-solid boundary, the last term 
in the right-hand side of Eq. (\ref{extrap-gnbc-int2}), 
$-\gamma\cos\theta_s^{surf}$, is the net wall force along $x$
arising from the wall-fluid interfacial free energy jump
across the fluid-fluid interface, in accordance with the Young's equation
$-\gamma\cos\theta_s^{surf}=\Delta\gamma_{wf}$.
On the other hand, the sum of the first three terms on the 
right-hand side of Eq. (\ref{extrap-gnbc-int2}) is the net fluid force 
along $x$ exerted on the three fluid sides of a BL fluid element in the 
interfacial region, due to the hydrodynamic motion of the fluids.

To obtain an extrapolated value for the contact angle $\theta_d^{surf}$ from 
Eq. (\ref{extrap-gnbc-int2}), we turn to the Stokes equation in the BL:
\begin{equation}\label{extrap-stokes}
-\partial_xp+\partial_x\sigma_{xx}^v+\partial_z\sigma_{zx}^v
+\mu\partial_x\phi=0,
\end{equation}
obtained from the $x$-component of Eq. (\ref{he1}) by dropping the
inertial and external forces.
Integration in $z$ of Eq. (\ref{extrap-stokes}) across the BL, together with 
the integration along $x$ across the fluid-fluid interface yields
\begin{equation}\label{extrap-stokes-int}
\Delta\left[\int_0^{z_0}dz\left(-p+\sigma_{xx}^v\right)\right]
+\int_{\rm int}dx\sigma_{zx}^v(z_0)+\gamma\cos\theta_d(z_0)
-\int_{\rm int}dx\sigma_{zx}^v(0)-\gamma\cos\theta_d^{surf}=0.
\end{equation}
Here we have made use of two relations: (1)
$\mu\partial_x\phi\simeq\gamma\kappa\delta(x-x_{\rm int})$ 
in the sharp interface limit \cite{sharp-interface-limit}, with
$\kappa$ being the interfacial curvature and $x_{\rm int}$ the
location of the interface along $x$. (2) $\int_0^{z_0}dz\kappa$
is the dynamic $z$-integrated curvature ${\cal K}_d=
\cos\theta_d(z_0)-\cos\theta_d^{surf}$. 
The local force balance along $x$ is expressed by Eq. (\ref{extrap-stokes}). 
Accordingly, the force balance along $x$ for the BL fluids in the 
integration region is expressed by Eq. (\ref{extrap-stokes-int}), where 
$\Delta\left[\int_0^{z_0}dz\left(-p+\sigma_{xx}^v\right)\right]$
is the net force on the left and right (constant-$x$) surfaces,
$\int_{\rm int}dx\sigma_{zx}^v(z_0)+\gamma\cos\theta_d(z_0)$
is the tangential force on the $z=z_0$ surface, and
$-\int_{\rm int}dx\sigma_{zx}^v(0)-\gamma\cos\theta_d^{surf}$ is the
tangential force on the $z=0$ surface.
Substituting Eq. (\ref{extrap-stokes-int}) into Eq. (\ref{extrap-gnbc-int2})
and identifying the normal stress $-p+\sigma_{xx}^v$ with $\sigma_{xx}^{HD}$,
we obtain 
\begin{equation}\label{extrap-gnbc-int3}
\int_{\rm int}dx\beta v^{\rm slip}_x=
\int_{\rm int}dx\sigma_{zx}^v(0)+\gamma\cos\theta_d^{surf}
-\gamma\cos\theta_s^{surf},
\end{equation}
which is identical to the integration of the continuum GNBC 
(Eq. (\ref{he3})) along $x$ across the fluid-fluid interface.

In summary, to obtain Eq. (\ref{extrap-gnbc-int3}) from
Eq. (\ref{extrap-gnbc-int1}), we have used both
$\partial_x\sigma_{xx}^{HS}+\partial_z\sigma_{zx}^0=0$ and
$\partial_x\sigma_{xx}^{HD}+\partial_z\sigma_{zx}=0$, whose
integrated expressions are given by Eq. (\ref{extrap-static})
and Eq. (\ref{extrap-stokes-int}), respectively.
We note that $\partial_x(\sigma_{xx}^{HD}-\sigma_{xx}^{HS})
+\partial_z(\sigma_{zx}-\sigma_{zx}^0)=0$ is equivalent to
the relation 
$\partial_x\tilde{\sigma}_{xx}+\partial_n\tilde{\sigma}_{nx}=0$
(integrated expressions given by Eqs. (\ref{angle1}) and (\ref{angle2})),
which has been used to obtain $\theta_d^{surf}$ through
extrapolation in Sec. \ref{md-evidence}.

\section{numerical algorithm}\label{app-numerical}
We present our numerical algorithm for solving the continuum
hydrodynamic model, comprising the dynamic equations
(\ref{he1}) and (\ref{he2}) and the four boundary conditions
$v_n=0$, $\partial_n\mu=0$, plus Eqs. (\ref{he3}) and (\ref{he4}).
We pay special attention to the application of boundary conditions,
and restrict our analysis to the Couette flow because 
the generalization to Poiseuille flow is straightforward.

\subsection{Dimensionless hydrodynamic equations}\label{numerical-dmleq}
To obtain a set of dimensionless equations suitable for 
numerical computations, we scale $\phi$ by 
$|\phi_\pm|=\sqrt{{r}/{u}}$, length by 
$\xi=\sqrt{{K}/{r}}$, velocity by the wall speed $V$, 
time by $\xi/V$, and pressure/stress by $\eta V/\xi$.
In dimensionless forms, the convection-diffusion equation reads
\begin{equation}\label{dmlcde}
\displaystyle\frac{\partial\phi}{\partial t}+
{\bf v}\cdot\nabla\phi={\cal L}_d\nabla^2(-\nabla^2\phi-\phi+\phi^3),
\end{equation}
the Navier-Stokes equation reads
\begin{equation}\label{dmlnse}
{\cal R}\left[\displaystyle\frac{\partial{\bf v}}{\partial t}+
\left({\bf v}\cdot\nabla\right){\bf v}\right]=-\nabla p+
\nabla^2{\bf v}+{\cal B}(-\nabla^2\phi-\phi+\phi^3 )\nabla\phi,
\end{equation}
the relaxation of $\phi$ at the fluid-solid boundary is governed by
\begin{equation}\label{dmlrelax}
\displaystyle\frac{\partial\phi}{\partial t}+v_x\partial_x\phi=
-{\cal V}_s\left[\partial_n\phi-
\displaystyle\frac{\sqrt{2}}{3}\cos\theta_s^0 s_\gamma(\phi)\right],
\end{equation}
and the GNBC becomes
\begin{equation}\label{dmlgnbc}
\left[{\cal L}_s(\phi)\right]^{-1} v^{slip}_x={\cal B}
\left[\partial_n\phi-\displaystyle\frac{\sqrt{2}}{3}\cos\theta_s^0 
s_\gamma(\phi)\right]\partial_x\phi-\partial_nv_x.
\end{equation}
Here $\beta(\phi)=(1-\phi)\beta_1/2+(1+\phi)\beta_2/2$
and $s_\gamma(\phi)=(\pi/2)\cos(\pi\phi/2)$.
Five dimensionless parameters appear in the above equations. 
They are (1) ${\cal L}_d={Mr}/{V\xi}$, 
which is the ratio of a diffusion length $Mr/V$ to $\xi$, 
(2) ${\cal R}={\rho V\xi}/{\eta}$,
(3) ${\cal B}={r^2\xi}/{u\eta V}={3\gamma}/{2\sqrt{2}\eta V}$,
which is inversely proportional to the capillary number $Ca=\eta V/\gamma$, 
(4) ${\cal V}_s={K\Gamma}/{V}$, which is the ratio of $K\Gamma$ 
(of velocity dimension) to $V$, 
and (5) ${\cal L}_s(\phi)={\eta}/{\beta(\phi)\xi}$, which is 
the ratio of the slip length $l_s(\phi)={\eta}/{\beta(\phi)}$ to $\xi$.

\subsection{Finite-difference scheme}\label{numerical-scheme}
For immiscible Couette flows, there are four variables 
$\phi$, $v_x$, $v_z$, and $p$ to be solved 
in a two-dimensional (2D) system (in the $xz$ plane). We want to solve 
the convection-diffusion equation and the Navier-Stokes equation 
in a 2D system of length $L_x$ (along $x$) and height $L_z$ (along $z$). 
Here $L_x$ must be large enough to allow the single phase flows (far from
the fluid-fluid interface) to approach uniform shear flows.
A finite-difference scheme is employed as follows.
(1) $N_x$ and $N_z$ equally spaced levels are introduced in the $x$ and $z$
directions, respectively. Grid size is given by $\Delta_x=L_x/(N_x-1)$
and $\Delta_z=L_z/(N_z-1)$ along $x$ and $z$, respectively.
(2) Each variable ($q$) is defined at $N_x\times N_z$ sites distributed from 
$x=-{L_x}/{2}$ to ${L_x}/{2}$ and from $z=-{L_z}/{2}$ to ${L_z}/{2}$,
represented by the array $q_{i,j}$, with $i=1,...,N_x$ and $j=1,...,N_z$.
Here $q_{i,j}\equiv q(x_i,z_j)$, with 
$x_i=(i-1)L_x/(N_x-1)-L_x/2$ and $z_j=(j-1)L_z/(N_z-1)-L_z/2$.
(3) In applying the various boundary conditions, ``ghost'' sites outside
the system, i.e., $i=0$, $i=N_x+1$, $j=0$, or $j=N_z+1$, may appear
in the discretization scheme. The values of the variables at the ghost sites
are determined separately from the various boundary conditions, detailed below. 
(4) First and second spatial derivatives along $\zeta$ (=$x$ or $z$)
are represented by
$\partial_\zeta q(\zeta_k)=[q(\zeta_{k+1})-q(\zeta_{k-1})]/{2\Delta_\zeta}$ 
and
$\partial^2_\zeta q(\zeta_k)=[q(\zeta_{k+1})+q(\zeta_{k-1})-2q(\zeta_{k})]
/{\Delta_\zeta^2}$.

\subsection{Convection-diffusion equation}\label{numerical-cde}
With the chemical potential $\mu_{i,j}$ given by
\begin{equation}\label{discmu}
\mu_{i,j}=-\left[
\displaystyle\frac{\phi_{i+1,j}-2\phi_{i,j}+\phi_{i-1,j}}{\Delta_x^2}+
\displaystyle\frac{\phi_{i,j+1}-2\phi_{i,j}+\phi_{i,j-1}}{\Delta_z^2}
\right]-\phi_{i,j}+\phi_{i,j}^3,
\end{equation}
the discretized convection-diffusion equation is
\begin{equation}\label{disccde}
\displaystyle\frac{\partial}{\partial t}\phi_{i,j}+
\left[{\bf v}\cdot\nabla\phi\right]_{i,j}={\cal L}_d
\left[\displaystyle\frac{\mu_{i+1,j}-2\mu_{i,j}+\mu_{i-1,j}}{\Delta_x^2}+
\displaystyle\frac{\mu_{i,j+1}-2\mu_{i,j}+\mu_{i,j-1}}{\Delta_z^2}\right],
\end{equation}
with
\begin{equation}\label{discconv}
\left[{\bf v}\cdot{\mbox{\boldmath$\nabla$}}\phi\right]_{i,j}=
v_{xi,j}\displaystyle\frac{\phi_{i+1,j}-\phi_{i-1,j}}{2\Delta_x}+  
v_{zi,j}\displaystyle\frac{\phi_{i,j+1}-\phi_{i,j-1}}{2\Delta_z}.
\end{equation}
The boundary conditions at $x=\pm L_x/2$ can be easily applied
using $\phi=\pm 1$ and 
\begin{equation}\label{single-couette}
{\bf v}(z)=\displaystyle\frac{L_z}{L_z+2L_s}
\displaystyle\frac{2z}{L_z}\hat{\bf x},
\end{equation}
for single-phase uniform shear flows. 
Here we focus on the boundary conditions at $z=\pm L_z/2$:
$\partial_n\mu=0$ and Eq. (\ref{dmlrelax}).
We spell out the numerics for the lower boundary $j=1$, with 
the understanding that the same can be applied to the upper boundary.

To solve the discretized convection-diffusion equation (\ref{disccde})
at the lower boundary $j=1$, we need the values of $\mu_{i,j}$ 
at $j=1$ and $j=0$. We also need the values of $\mu_{i,j}$ at $j=1$ 
to solve the same equation at $j=2$. According to Eq. (\ref{discmu}),
$\mu_{i,j}$ at $j=1$ and $j=0$ can not be directly evaluated from
$\phi_{i,j}$ with $i=1,...,N_x$ and $j=1,...,N_z$. But they can still 
be determined from the boundary conditions at $z=-L_z/2$.
$\mu_{i,j}$ at $j=0$ is obtained from $\partial_n\mu=0$ at $j=1$ as
\begin{equation}\label{cdebc1}
\mu_{i,j-1=0}=\mu_{i,j+1=2}.
\end{equation}
To obtain $\mu_{i,j}$ at $j=1$, we need to determine $\phi_{i,j}$ at $j=0$.
This can be done by requiring that Eqs. (\ref{dmlcde}) and (\ref{dmlrelax})
yield the same $\partial\phi/\partial t$ at $z=-L_z/2$.
The discretized convection-diffusion equation is given by Eq. (\ref{disccde})
while the discretized relaxation equation for $\phi$ at the boundary
$j=1$ is given by
\begin{equation}\label{cdebc2}
\displaystyle\frac{\partial}{\partial t}\phi_{i,j}+
\left[{\bf v}\cdot\nabla\phi\right]_{i,j}
=-{\cal V}_s\left[
\displaystyle\frac{\phi_{i,j-1}-\phi_{i,j+1}}{2\Delta_z}-
\displaystyle\frac{\sqrt{2}}{3}\cos\theta_s^0 s_\gamma(\phi_{i,j})
\right].
\end{equation}
Equating the right-hand side of Eq. (\ref{disccde}) at $j=1$
(with $\mu_{i,0}$ fixed by Eq. (\ref{cdebc1}) and other $\mu$'s 
given by Eq. (\ref{discmu})) with that of Eq. (\ref{cdebc2}) leads to 
a tridiagonal system of linear equations for $\phi_{i,j}$ ($\phi_{i,j}$
coupled with $\phi_{i-1,j}$ and $\phi_{i+1,j}$) at $j=0$. Solving this
tridiagonal system determines $\phi_{i,j}$ at $j=0$, from which we
obtain $\mu_{i,j}$ at $j=1$ by using Eq. (\ref{discmu}).

\subsection{Navier-Stokes equation}\label{numerical-nse}
We now turn to the Navier-Stokes equation (\ref{dmlnse})
with the incompressibility condition $\nabla\cdot{\bf v}=0$.
The difficulty in solving the Navier-Stokes equation 
is the lack of a time evolution equation for the pressure $p$. 
In the following, we will introduce a numerical method 
based on the pressure Poisson equation \cite{liu}.

\subsubsection{Pressure Poisson equation}\label{numerical-nse-ppe}
Taking the divergence of momentum equation (\ref{dmlnse}) 
and applying the incompressibility condition, 
we obtain the pressure Poisson equation
\begin{equation}\label{ppe}
\nabla^2p=-{\cal R}\nabla\cdot\left[
\left({\bf v}\cdot\nabla\right){\bf v}\right]
+{\cal B}\nabla\cdot[(-\nabla^2\phi-\phi+\phi^3)\nabla\phi].
\end{equation}
Dotting the momentum equation (\ref{dmlnse}) with the surface normal 
at the fluid-solid boundary and using $v_n=0$, 
we obtain for Eq. (\ref{ppe}) the boundary condition
\begin{equation}\label{ppebc}
\partial_n p=\nabla^2v_n+{\cal B}
(-\nabla^2\phi-\phi+\phi^3)\partial_n\phi,
\end{equation}
at $z=\pm L_z/2$. 
In addition, we use $\nabla^2p=0$ and $\partial_xp=0$ for
the values of $\nabla^2 p$ and $\partial_n p$ at the boundaries $x=\pm L_x/2$.
This reflects the single phase flow given by Eq. (\ref{single-couette}).

From the momentum equation (\ref{dmlnse}) 
and the pressure Poisson equation (\ref{ppe}), 
we derive a diffusion equation 
$$
{\cal R}\displaystyle\frac{\partial(\nabla\cdot{\bf v})}{\partial t}
=\nabla^2(\nabla\cdot{\bf v}),
$$
for $\nabla\cdot{\bf v}$.
With $\nabla\cdot{\bf v}=0$ given at time $t=0$, and
in order to ensure that ${\bf v}$ remains divergence-free at $t>0$, 
we must impose the additional boundary condition 
$\nabla\cdot{\bf v}=0$ at all times $t\ge 0$.
We will show that this boundary condition is needed 
in solving for $p$ in a finite-difference scheme.

In order to solve the pressure Poisson equation, we need to evaluate
$[\nabla^2p]_{i,j}$ for $i=1,...,N_x$ and $j=1,...,N_z$,
$[\partial_xp]_{i,j}$ for $i=1,N_x$ and $j=1,...,N_z$, and
$[\partial_zp]_{i,j}$ for $i=1,...,N_x$ and $j=1,N_z$.
For $\nabla^2p$, we have
$$
[\nabla^2p]_{i,j}=0,
$$
for $i=1,N_x$ and $j=1,...,N_z$;
$$\begin{array}{ll}
[\nabla^2p]_{i,j}=& 2{\cal R}\left[
\displaystyle\frac{v_{xi+1,j}-v_{xi-1,j}}{2\Delta_x}
\displaystyle\frac{v_{zi,j+1}-v_{zi,j-1}}{2\Delta_z}
-\displaystyle\frac{v_{zi+1,j}-v_{zi-1,j}}{2\Delta_x}
\displaystyle\frac{v_{xi,j+1}-v_{xi,j-1}}{2\Delta_z}\right] \\&
+{\cal B}\mu_{i,j}\left(
\displaystyle\frac{\phi_{i+1,j}-2\phi_{i,j}+\phi_{i-1,j}}{\Delta_x^2}+
\displaystyle\frac{\phi_{i,j+1}-2\phi_{i,j}+\phi_{i,j-1}}{\Delta_z^2}\right)\\&
+{\cal B}\left(\displaystyle\frac{\mu_{i+1,j}-\mu_{i-1,j}}{2\Delta_x}
\displaystyle\frac{\phi_{i+1,j}-\phi_{i-1,j}}{2\Delta_x}
+\displaystyle\frac{\mu_{i,j+1}-\mu_{i,j-1}}{2\Delta_z}
\displaystyle\frac{\phi_{i,j+1}-\phi_{i,j-1}}{2\Delta_z}\right),
\end{array}$$
for $i=2,...,N_x-1$ and $j=2,...,N_z-1$; and
$$\begin{array}{ll}
[\nabla^2p]_{i,j}=& 2{\cal R}
\displaystyle\frac{v_{xi+1,j}-v_{xi-1,j}}{2\Delta_x}
\displaystyle\frac{v_{zi,j+1}-v_{zi,j-1}}{2\Delta_z} \\&
+{\cal B}\mu_{i,j}\left(
\displaystyle\frac{\phi_{i+1,j}-2\phi_{i,j}+\phi_{i-1,j}}{\Delta_x^2}+
\displaystyle\frac{\phi_{i,j+1}-2\phi_{i,j}+\phi_{i,j-1}}{\Delta_z^2}\right)\\&
+{\cal B}\displaystyle\frac{\mu_{i+1,j}-\mu_{i-1,j}}{2\Delta_x}
\displaystyle\frac{\phi_{i+1,j}-\phi_{i-1,j}}{2\Delta_x},
\end{array}$$
for $i=2,...,N_x-1$ and $j=1,N_z$ (where $v_z=0$ and $\partial_z\mu=0$). 
We see that $\phi$ and $v_z$ 
at ghost sites of $j=0,N_z+1$ appear in the last expression. 
The ghost $\phi$'s have already been determined 
in solving the convection-diffusion equation, while the ghost
$v_z$'s are determined through the additional boundary condition
$\nabla\cdot{\bf v}=0$:
$$
\displaystyle\frac{v_{xi+1,j}-v_{xi-1,j}}{2\Delta_x}+
\displaystyle\frac{v_{zi,j+1}-v_{zi,j-1}}{2\Delta_z}=0, 
$$
for $i=2,...,N_x-1$, and $j=1,N_z$.
For $\partial_np$, we have 
$$[\partial_xp]_{i,j}=0$$ for $i=1,N_x$ and $j=1,...,N_z$; 
$$[\partial_zp]_{i,j}=0$$ for $i=1,N_x$ and $j=1,N_z$; and
$$
[\partial_zp]_{i,j}=
\displaystyle\frac{v_{zi,j+1}+v_{zi,j-1}}{\Delta_z^2}+{\cal B}
\mu_{i,j}\displaystyle\frac{\phi_{i,j+1}-\phi_{i,j-1}}{2\Delta_z}
$$
for $i=2,...,N_x-1$ and $j=1,N_z$ (where $v_z=0$).
The last expression involves the ghost $\phi$'s  and
$v_z$'s at $j=0,N_z+1$.
Given the above values of $[\nabla^2p]_{i,j}$ and $[\partial_n p]_{i,j}$,
we apply a 2D Fast Fourier Transformation to solve $p_{i,j}(0)$ 
(up to a constant) for $i=1,...,N_x$ and $j=1,...,N_z$.

\subsubsection{Slip boundary condition}\label{numerical-nse-slip}
The discretized Navier-Stokes equation is given by
\begin{equation}\label{discnsex}
\begin{array}{ll}
\displaystyle\frac{\partial v_{xi,j}}{\partial t}= &
-v_{xi,j}\displaystyle\frac{v_{xi+1,j}-v_{xi-1,j}}{2\Delta_x}
-v_{zi,j}\displaystyle\frac{v_{xi,j+1}-v_{xi,j-1}}{2\Delta_z}
-\displaystyle\frac{1}{\cal R}
\displaystyle\frac{p_{i+1,j}-p_{i-1,j}}{2\Delta_x} \\&
+\displaystyle\frac{1}{\cal R}\left( \displaystyle\frac
{v_{xi+1,j}-2v_{xi,j}+v_{xi-1,j}}{\Delta_x^2}+\displaystyle\frac
{v_{xi,j+1}-2v_{xi,j}+v_{xi,j-1}}{\Delta_z^2} \right) \\&
+\displaystyle\frac{\cal B}{\cal R}\mu_{i,j}
\displaystyle\frac{\phi_{i+1,j}-\phi_{i-1,j}}{2\Delta_x},
\end{array}
\end{equation}
for $i=2,...,N_x-1$ and $j=1,...,N_z$, and
\begin{equation}\label{discnsez}
\begin{array}{ll}
\displaystyle\frac{\partial v_{zi,j}}{\partial t}= &
-v_{xi,j}\displaystyle\frac{v_{zi+1,j}-v_{zi-1,j}}{2\Delta_x}
-v_{zi,j}\displaystyle\frac{v_{zi,j+1}-v_{zi,j-1}}{2\Delta_z}
-\displaystyle\frac{1}{\cal R}
\displaystyle\frac{p_{i,j+1}-p_{i,j-1}}{2\Delta_z} \\&
+\displaystyle\frac{1}{\cal R}\left( \displaystyle\frac
{v_{zi+1,j}-2v_{zi,j}+v_{zi-1,j}}{\Delta_x^2}+\displaystyle\frac
{v_{zi,j+1}-2v_{zi,j}+v_{zi,j-1}}{\Delta_z^2} \right) \\&
+\displaystyle\frac{\cal B}{\cal R}\mu_{i,j}
\displaystyle\frac{\phi_{i,j+1}-\phi_{i,j-1}}{2\Delta_z}.
\end{array}
\end{equation}
for $i=2,...,N_x-1$ and $j=2,...,N_z-1$, together with 
the boundary conditions that $v_{zi,j}=0$ at $j=1,N_z$
and $\bf v$ given by Eq. (\ref{single-couette}) at $i=1,N_x$.
Equation (\ref{discnsex}) at $j=1,N_z$ involves $\phi$ and $v_x$ 
at ghost sites of $j=0,N_z+1$. 
The ghost $\phi$'s come from $\mu_{i,j}$ at $j=1,N_z$, and
have already been determined. The ghost $v_x$'s are determined
from the discretized GNBC
\begin{equation}\label{discgnbcl}
\left[{\cal L}_s(\phi_{i,j})\right]^{-1}v_{xi,j}^{slip}=
{\cal B}\left[\displaystyle\frac{\phi_{i,j-1}-\phi_{i,j+1}}
{2\Delta_z}-
\displaystyle\frac{\sqrt{2}}{3}\cos\theta_s^0 s_\gamma(\phi_{i,j})
\right]\displaystyle\frac{\phi_{i+1,j}-\phi_{i-1,j}}{2\Delta_x}
-\displaystyle\frac{v_{xi,j-1}-v_{xi,j+1}}{2\Delta_z},
\end{equation}
at the lower boundary $j=1$ with $v_{xi,j}^{slip}=v_{xi,j}-V$, and
\begin{equation}\label{discgnbcu}
\left[{\cal L}_s(\phi_{i,j})\right]^{-1}v_{xi,j}^{slip}=
{\cal B}\left[\displaystyle\frac{\phi_{i,j+1}-\phi_{i,j-1}}
{2\Delta_z}-
\displaystyle\frac{\sqrt{2}}{3}\cos\theta_s^0 s_\gamma(\phi_{i,j})
\right]\displaystyle\frac{\phi_{i+1,j}-\phi_{i-1,j}}{2\Delta_x}
-\displaystyle\frac{v_{xi,j+1}-v_{xi,j-1}}{2\Delta_z},
\end{equation}
at the upper boundary $j=N_z$ with $v_{xi,j}^{slip}=v_{xi,j}+V$.

In summary, to solve the dynamic equations (\ref{he1}) and (\ref{he2}),
we need to use $\phi=\pm 1$ and Eq. (\ref{single-couette}) 
at $x=\pm L_x/2$, with $v_n=0$, $\partial_n\mu=0$, plus 
Eqs. (\ref{he3}) and (\ref{he4}) at $z=\pm L_z/2$. In particular,
in applying the boundary conditions at $z=\pm L_z/2$, values of $\phi$,
$v_x$, and $v_z$ at ghost sites have to be introduced and solved for.

\subsection{Time integration}
We outline the scheme for time discretization and integration. 
For simplicity we only describe the forward Euler time stepping.
In the following a superscript $n$ denotes consecutive time instants 
and $\Delta t$ is the time interval.

\parindent=0pt
{\bf Time Stepping}: Given $\left\{\phi_{i,j}^n\right\}$ and
$\left\{{\bf v}_{i,j}^n\right\}$ at all the sites 
($i=1,...,N_x$ and $j=1,...,N_z$) in the system:

{\bf Step 1}: Determine $\left\{\mu_{i,j}^n\right\}$,
$\left\{\phi_{i,j}^n\right\}$, and
$\left\{{\bf v}_{i,j}^n\right\}$ at the ghost sites 
from the various boundary conditions, as described in Secs. 
\ref{numerical-cde}, \ref{numerical-nse-ppe}, and \ref{numerical-nse-slip}.

{\bf Step 2}: Solve $\left\{p_{i,j}^n\right\}$ at all the interior sites
($i=1,...,N_x$ and $j=1,...,N_z$) from Eq. (\ref{ppe}) with appropriate
boundary conditions for $\partial_n p$, as described in 
Sec. \ref{numerical-nse-ppe}.

{\bf Step 3}: Compute $\left\{\phi_{i,j}^{n+1}\right\}$ and
$\left\{{\bf v}_{i,j}^{n+1}\right\}$ at all the interior sites 
(except those fixed by the boundary conditions at all times) using
$$
\displaystyle\frac{\phi^{n+1}-\phi^n}{\Delta t}=
-{\bf v}^n\cdot\nabla\phi^n+{\cal L}_d\nabla^2\mu^n,
$$
and
$$
{\cal R}\displaystyle\frac{{\bf v}^{n+1}-{\bf v}^n}{\Delta t}=
-{\cal R}\left({\bf v}^n\cdot\nabla\right){\bf v}^n-\nabla p^n+
\nabla^2{\bf v}^n+{\cal B}\mu^n\nabla\phi^n,
$$
according to Eqs. (\ref{disccde}), (\ref{discnsex}), and
(\ref{discnsez}) in discretized time. Here the ghost 
$\left\{\mu_{i,j}^n\right\}$, $\left\{\phi_{i,j}^n\right\}$,
and $\left\{{\bf v}_{i,j}^n\right\}$ determined in Step 1 and 
$\left\{p_{i,j}^n\right\}$ solved in Step 2 are needed.

\begin{references}
\bibitem{balance}
Inertial effect was found to be less than $\sigma_{nx}$
by two orders of magnitude.
\bibitem{weight} This relation is obtained by assuming the two
fluids interacting independently with the wall, so that
$\tilde{G}_x^f$  may be expressed as the weighted average of
$\tilde{G}_x^{f1}=\beta_1 v_x^{slip1}$ and
$\tilde{G}_x^{f2}=\beta_2 v_x^{slip2}$.
The desired expression is obtained by noting
$v_x^{slip1}\simeq v_x^{slip2}$ to within $10\%$.
\bibitem{surface-angle}
For gently curved interfaces,
$\int_{\rm int}dx[K\partial_n\phi\partial_x\phi]=
\int_{\rm int}d\phi(K\partial_l\phi)\cos\theta_d^{surf}$,
where $\partial_l$ means taking partial derivative with respect to
(fluid-fluid) interface normal $l$, with $\partial_n\phi$ 
approximated by $\partial_l\phi\cos\theta_d^{surf}$. As
$\int_{\rm int}d\phi(K\partial_l\phi)=
\int_{\rm int}dl K(\partial_l\phi)^2=\gamma$, we have
$\int_{\rm int}dx[K\partial_n\phi\partial_x\phi]=\gamma\cos\theta_d^{surf}$.
\bibitem{bc}
Agreement between continuum hydrodynamics and MD simulation
has been demonstrated by treating the slip profile 
($v_x^{slip}$ along the wall, derived from MD simulations)
and the dynamic contact angle as inputs to the continuum model,
see Ref. \cite{hadji}.
The significance of our results lies in the fact that no such
information have been put in by hand.
\bibitem{breakup}
The breakup of the interface (with a film left behind the moving CL)
has been observed for both the Couette and Poiseuille flows
at large $Ca$ \cite{koplik,robbins}.
\end{references}
\end{document}